\pdfoutput=1
\documentclass[12pt,a4paper]{article}

\usepackage{ifthen} 
\newboolean{pdflatex}
\setboolean{pdflatex}{true} 

\newboolean{articletitles}
\setboolean{articletitles}{true} 

\newboolean{uprightparticles}
\setboolean{uprightparticles}{false} 

\def\paperauthors{LHCb collaboration} 
\def\paperasciititle{Measurement of gamma using B->DK and B->Dpi decays with D->Kspipi and D->KsKK} 
\def\papertitle{Measurement of \g using \BtoDK and \BtoDpi decays with \DtoKspp and \DtoKskk} 
\def\paperkeywords{{High Energy Physics}, {LHCb}} 
\def\papercopyright{\the\year\ CERN for the benefit of the LHCb collaboration} 
\def\paperlicence{CC BY 4.0 licence}
\def\paperlicenceurl{https://creativecommons.org/licenses/by/4.0/}

\newif\ifEnableSectionTOCLinks
\EnableSectionTOCLinksfalse 

\usepackage[top=1in, bottom=1.25in, left=1in, right=1in]{geometry}

\usepackage{microtype}
\usepackage{lineno}  
\usepackage{xspace} 
\usepackage{caption} 

\usepackage{graphicx}  
\usepackage{color}
\usepackage{colortbl}
\graphicspath{{./figs/}} 

\usepackage{amsmath} 
\usepackage{amssymb}
\usepackage{amsfonts}
\usepackage{upgreek} 

\newcommand*\patchAmsMathEnvironmentForLineno[1]{%
\expandafter\let\csname old#1\expandafter\endcsname\csname #1\endcsname
\expandafter\let\csname oldend#1\expandafter\endcsname\csname
end#1\endcsname
 \renewenvironment{#1}%
   {\linenomath\csname old#1\endcsname}%
   {\csname oldend#1\endcsname\endlinenomath}%
}
\newcommand*\patchBothAmsMathEnvironmentsForLineno[1]{%
  \patchAmsMathEnvironmentForLineno{#1}%
  \patchAmsMathEnvironmentForLineno{#1*}%
}
\AtBeginDocument{%
\patchBothAmsMathEnvironmentsForLineno{equation}%
\patchBothAmsMathEnvironmentsForLineno{align}%
\patchBothAmsMathEnvironmentsForLineno{flalign}%
\patchBothAmsMathEnvironmentsForLineno{alignat}%
\patchBothAmsMathEnvironmentsForLineno{gather}%
\patchBothAmsMathEnvironmentsForLineno{multline}%
\patchBothAmsMathEnvironmentsForLineno{eqnarray}%
}

\usepackage[pdftex,
            pdfauthor={\paperauthors},
            pdftitle={\paperasciititle},
            pdfkeywords={\paperkeywords}]{hyperref}
\usepackage{hyperxmp}
\hypersetup{
    pdfcopyright={Copyright (C) \papercopyright},
    pdflicenseurl={\paperlicenceurl}
}

\usepackage[colorinlistoftodos,textsize=scriptsize]{todonotes}

\usepackage[bottom,flushmargin,hang,multiple]{footmisc}

\usepackage[all]{hypcap} 

\usepackage{xspace} 
\usepackage{upgreek}

\def\lhcb   {\mbox{LHCb}\xspace}

\def\besiii {\mbox{BESIII}\xspace}
\def\cleo   {\mbox{CLEO}\xspace}

\def\MagUp {\mbox{\em Mag\kern -0.05em Up}\xspace}

\ifthenelse{\boolean{uprightparticles}}%
{

 \def\Pmu         {\ensuremath{\upmu}\xspace}

 \def\Ppi         {\ensuremath{\uppi}\xspace}

 \def\PDelta      {\ensuremath{\Delta}\xspace}                 
 \def\PXi         {\ensuremath{\Xi}\xspace}                 
 \def\PLambda     {\ensuremath{\Lambda}\xspace}                 
 \def\PSigma      {\ensuremath{\Sigma}\xspace}                 
 \def\POmega      {\ensuremath{\Omega}\xspace}                 
 \def\PUpsilon    {\ensuremath{\Upsilon}\xspace}
 \let\oldPi\Pi
 \def\PPi         {\ensuremath{\oldPi}\xspace}

 \def\PB      {\ensuremath{\mathrm{B}}\xspace}                 
 \def\PD      {\ensuremath{\mathrm{D}}\xspace}                 

 \def\PK      {\ensuremath{\mathrm{K}}\xspace}                 
 \def\Pb      {\ensuremath{\mathrm{b}}\xspace}                 
 \def\Pc      {\ensuremath{\mathrm{c}}\xspace}                 
 \def\Pd      {\ensuremath{\mathrm{d}}\xspace}

 \def\Ph      {\ensuremath{\mathrm{h}}\xspace}                 
 \def\Ps      {\ensuremath{\mathrm{s}}\xspace}                 
                  
 \def\Pu      {\ensuremath{\mathrm{u}}\xspace}                 
 \def\thebaroffset{0.0em}
}
{

 \def\Pmu         {\ensuremath{\mu}\xspace}

 \def\Ppi         {\ensuremath{\pi}\xspace}

 \mathchardef\PDelta="7101
 \mathchardef\PXi="7104
 \mathchardef\PLambda="7103
 \mathchardef\PSigma="7106
 \mathchardef\POmega="710A
 \mathchardef\PUpsilon="7107
 \mathchardef\PPi="7105
 \def\PB      {\ensuremath{B}\xspace}                 
 \def\PD      {\ensuremath{D}\xspace}                 

 \def\PK      {\ensuremath{K}\xspace}                 
 \def\Pb      {\ensuremath{b}\xspace}                 
 \def\Pc      {\ensuremath{c}\xspace}                 
 \def\Pd      {\ensuremath{d}\xspace}

 \def\Ph      {\ensuremath{h}\xspace}                 
 \def\Ps      {\ensuremath{s}\xspace}                 
                  
 \def\Pu      {\ensuremath{u}\xspace}                 
 \def\thebaroffset{0.18em}
}
\newcommand{\offsetoverline}[2][\thebaroffset]{\kern #1\overline{\kern -#1 #2}}%

\makeatletter
\ifcase \@ptsize \relax
  \newcommand{\miniscule}{\@setfontsize\miniscule{4}{5}}
\or
  \newcommand{\miniscule}{\@setfontsize\miniscule{5}{6}}
\or
  \newcommand{\miniscule}{\@setfontsize\miniscule{5}{6}}
\fi
\makeatother

\DeclareRobustCommand{\optbar}[1]{\shortstack{{\miniscule (\rule[.5ex]{1.25em}{.18mm})}
  \\ [-.7ex] $#1$}}

\def\mup        {{\ensuremath{\Pmu^+}}\xspace}

\def\g      {{\ensuremath{\Pgamma}}\xspace}

\def\uquark    {{\ensuremath{\Pu}}\xspace}
\def\uquarkbar {{\ensuremath{\overline \uquark}}\xspace}

\def\dquark    {{\ensuremath{\Pd}}\xspace}

\def\squark    {{\ensuremath{\Ps}}\xspace}
\def\squarkbar {{\ensuremath{\overline \squark}}\xspace}

\def\cquark    {{\ensuremath{\Pc}}\xspace}
\def\cquarkbar {{\ensuremath{\overline \cquark}}\xspace}

\def\bquark    {{\ensuremath{\Pb}}\xspace}
\def\bquarkbar {{\ensuremath{\overline \bquark}}\xspace}

\def\hadron {{\ensuremath{\Ph}}\xspace}
\def\pion   {{\ensuremath{\Ppi}}\xspace}
\def\piz    {{\ensuremath{\pion^0}}\xspace}
\def\pip    {{\ensuremath{\pion^+}}\xspace}
\def\pim    {{\ensuremath{\pion^-}}\xspace}
\def\pipm   {{\ensuremath{\pion^\pm}}\xspace}
\def\pimp   {{\ensuremath{\pion^\mp}}\xspace}

\def\kaon    {{\ensuremath{\PK}}\xspace}

\def\KorKbar {\kern \thebaroffset\optbar{\kern -\thebaroffset \PK}{}\xspace}

\def\Kp      {{\ensuremath{\kaon^+}}\xspace}
\def\Km      {{\ensuremath{\kaon^-}}\xspace}
\def\Kpm     {{\ensuremath{\kaon^\pm}}\xspace}
\def\Kmp     {{\ensuremath{\kaon^\mp}}\xspace}
\def\KS      {{\ensuremath{\kaon^0_{\mathrm{S}}}}\xspace}

\def\KL      {{\ensuremath{\kaon^0_{\mathrm{L}}}}\xspace}
\def\Kstarz  {{\ensuremath{\kaon^{*0}}}\xspace}

\def\Kstarp  {{\ensuremath{\kaon^{*+}}}\xspace}

\def\Kstarpm {{\ensuremath{\kaon^{*\pm}}}\xspace}

\def\Dbar    {{\ensuremath{\offsetoverline{\PD}}}\xspace}
\def\D       {{\ensuremath{\PD}}\xspace}

\def\DorDbar {\kern \thebaroffset\optbar{\kern -\thebaroffset \PD}\xspace}
\def\Dz      {{\ensuremath{\D^0}}\xspace}
\def\Dzb     {{\ensuremath{\Dbar{}^0}}\xspace}
\def\Dp      {{\ensuremath{\D^+}}\xspace}
\def\Dm      {{\ensuremath{\D^-}}\xspace}

\def\DpDm    {\ensuremath{\Dp {\kern -0.16em \Dm}}\xspace}
\def\Dstar   {{\ensuremath{\D^*}}\xspace}

\def\Dstarz  {{\ensuremath{\D^{*0}}}\xspace}

\def\Dstarpm {{\ensuremath{\D^{*\pm}}}\xspace}

\def\B       {{\ensuremath{\PB}}\xspace}

\def\BorBbar {\kern \thebaroffset\optbar{\kern -\thebaroffset \PB}\xspace}
\def\Bz      {{\ensuremath{\B^0}}\xspace}

\def\Bd      {{\ensuremath{\B^0}}\xspace}

\def\BdorBdbar {\kern \thebaroffset\optbar{\kern -\thebaroffset \Bd}\xspace}
\def\Bu      {{\ensuremath{\B^+}}\xspace}
\def\Bub     {{\ensuremath{\B^-}}\xspace}
\def\Bp      {{\ensuremath{\Bu}}\xspace}
\def\Bm      {{\ensuremath{\Bub}}\xspace}
\def\Bpm     {{\ensuremath{\B^\pm}}\xspace}

\def\Bs      {{\ensuremath{\B^0_\squark}}\xspace}

\def\BsorBsbar {\kern \thebaroffset\optbar{\kern -\thebaroffset \Bs}\xspace}

\def\Y#1S{\ensuremath{\PUpsilon{(#1S)}}\xspace}

\def\Lz          {{\ensuremath{\PLambda}}\xspace}

\def\LorLbar     {\kern \thebaroffset\optbar{\kern -\thebaroffset \PLambda}\xspace}

\def\Lb           {{\ensuremath{\Lz^0_\bquark}}\xspace}

\newcommand{\decay}[2]{\ensuremath{\mathinner{#1\!\to #2}}\xspace}

\def\to                 {\ensuremath{\rightarrow}\xspace}

\def\CP                {{\ensuremath{C\!P}}\xspace}

\def\Vud  {{\ensuremath{V_{\uquark\dquark}^{\phantom{\ast}}}}\xspace}
\def\Vcd  {{\ensuremath{V_{\cquark\dquark}^{\phantom{\ast}}}}\xspace}

\def\Vubs  {{\ensuremath{V_{\uquark\bquark}^\ast}}\xspace}
\def\Vcbs  {{\ensuremath{V_{\cquark\bquark}^\ast}}\xspace}

\def\AT#1     {\ensuremath{A_{\mathrm{T}}^{#1}}\xspace}           

\def\C#1      {\ensuremath{\mathcal{C}_{#1}}\xspace}                       
\def\Cp#1     {\ensuremath{\mathcal{C}_{#1}^{'}}\xspace}                    
\def\Ceff#1   {\ensuremath{\mathcal{C}_{#1}^{\mathrm{(eff)}}}\xspace}        
\def\Cpeff#1  {\ensuremath{\mathcal{C}_{#1}^{'\mathrm{(eff)}}}\xspace}       
\def\Ope#1    {\ensuremath{\mathcal{O}_{#1}}\xspace}                       
\def\Opep#1   {\ensuremath{\mathcal{O}_{#1}^{'}}\xspace}                    

\newcommand{\aunit}[1]{\ensuremath{\text{\,#1}}}       

\newcommand{\tev}{\aunit{Te\kern -0.1em V}\xspace}
\newcommand{\gev}{\aunit{Ge\kern -0.1em V}\xspace}
\newcommand{\mev}{\aunit{Me\kern -0.1em V}\xspace}
\newcommand{\kev}{\aunit{ke\kern -0.1em V}\xspace}
\newcommand{\ev}{\aunit{e\kern -0.1em V}\xspace}
 
\newcommand{\mevc}{\ensuremath{\aunit{Me\kern -0.1em V\!/}c}\xspace}
\newcommand{\gevc}{\ensuremath{\aunit{Ge\kern -0.1em V\!/}c}\xspace}
\newcommand{\mevcc}{\ensuremath{\aunit{Me\kern -0.1em V\!/}c^2}\xspace}
\newcommand{\gevcc}{\ensuremath{\aunit{Ge\kern -0.1em V\!/}c^2}\xspace}

\def\fb   {\ensuremath{\aunit{fb}}\xspace}
\def\invfb   {\ensuremath{\fb^{-1}}\xspace}

\def\ci {\aunit{Ci}\xspace}

\newcommand{\chisq}{\ensuremath{\chi^2}\xspace}

\newcommand{\chisqip}{\ensuremath{\chi^2_{\text{IP}}}\xspace}

\def\deriv {\ensuremath{\mathrm{d}}}

\def\gsim{{~\raise.15em\hbox{$>$}\kern-.85em
          \lower.35em\hbox{$\sim$}~}\xspace}
\def\lsim{{~\raise.15em\hbox{$<$}\kern-.85em
          \lower.35em\hbox{$\sim$}~}\xspace}

\def\PDF {PDF\xspace}

\def\sPlot{\mbox{\em sPlot}\xspace}

\def\sqs   {\ensuremath{\protect\sqrt{s}}\xspace}

\def\evtgen     {\mbox{\textsc{EvtGen}}\xspace}

\def\geant      {\mbox{\textsc{Geant4}}\xspace}

\def\photos     {\mbox{\textsc{Photos}}\xspace}

\def\pythia     {\mbox{\textsc{Pythia}}\xspace}

\def\tell1  {TELL1\xspace}
\def\ukl1   {UKL1\xspace}

\newcommand{\vs}{\mbox{\itshape vs.}\xspace}

\newcommand{\lhcborcid}[1]{\href{https://orcid.org/#1}{\hspace*{0.1em}\raisebox{-0.45ex}{\includegraphics[width=1em]{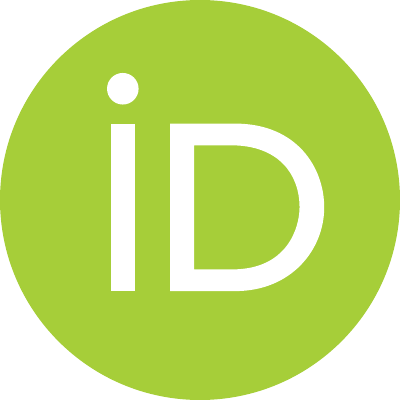}}}}

\newcommand{\Kspipi}{\ensuremath{\KS\pip\pim}\xspace}

\newcommand{\KsKK}{\ensuremath{\KS\Kp\Km}\xspace}

\newcommand{\DtoKsKK}{\ensuremath{\PD\to\KsKK}\xspace}
\newcommand{\DtoKshh}{\ensuremath{\PD\to\Kshh}\xspace}

\newcommand{\BtoDpi}{\ensuremath{\Bpm\to\PD\pipm}\xspace}
\newcommand{\BtoDK}{\ensuremath{\Bpm\to\PD\Kpm}\xspace}
\newcommand{\BtoDh}{\ensuremath{\PB\to\PD\Ph}\xspace}

\newcommand{\Dpi}{\ensuremath{\PD\pi^\pm}\xspace}

\newcommand{\DK}{\ensuremath{\PD\PK^\pm}\xspace}

\newcommand{\msqmin}{\ensuremath{m^2_-}\xspace}
\newcommand{\msqplus}{\ensuremath{m^2_+}\xspace}

\def        \BtoDpi         {\mbox{\ensuremath{{\Bpm\to\D\pipm}}}\xspace}
\def        \BtoDK          {\mbox{\ensuremath{{\Bpm\to\D\Kpm}}}\xspace}

\def        \DK             {{\D\Kpm}\xspace}
\def        \BtoDh          {\ensuremath{\Bpm\to\D\hadron^\pm}\xspace}
\def        \Kspp           {\mbox{\ensuremath{\KS\pip\pim}}\xspace}
\def        \Kskk           {\mbox{\ensuremath{\KS\Kp\Km}}\xspace}
\def        \DtoKspp        {\mbox{\ensuremath{\D\to\KS\pip\pim}}\xspace}
\def        \DtoKskk        {\mbox{\ensuremath{\D\to\KS\Kp\Km}}\xspace}

\def        \DtoKshh        {\ensuremath{\D\to\KS\hadron^+\hadron^-}\xspace}

\newcommand{\xm}{\ensuremath{x_-}\xspace}
\newcommand{\ym}{\ensuremath{y_-}\xspace}
\newcommand{\xp}{\ensuremath{x_+}\xspace}
\newcommand{\yp}{\ensuremath{y_+}\xspace}

\newcommand{\xpmdk}{\ensuremath{x_\pm^{\D\kaon}}\xspace}
\newcommand{\ypmdk}{\ensuremath{y_\pm^{\D\kaon}}\xspace}
\newcommand{\xpmdpi}{\ensuremath{x_\pm^{\D\pi}}\xspace}
\newcommand{\ypmdpi}{\ensuremath{y_\pm^{\D\pi}}\xspace}
\newcommand{\xmdk}{\ensuremath{x_-^{\D\kaon}}\xspace}
\newcommand{\ymdk}{\ensuremath{y_-^{\D\kaon}}\xspace}
\newcommand{\xpdk}{\ensuremath{x_+^{\D\kaon}}\xspace}
\newcommand{\ypdk}{\ensuremath{y_+^{\D\kaon}}\xspace}
\newcommand{\xxidpi}{\ensuremath{x_\xi^{\D\pi}}\xspace}
\newcommand{\yxidpi}{\ensuremath{y_\xi^{\D\pi}}\xspace}

\renewcommand{\Re}{\text{Re}}
\renewcommand{\Im}{\text{Im}}

\newcommand{\dpi}{\ensuremath{\D\pi}\xspace}

\renewcommand{\g}{\ensuremath{\gamma}\xspace}
\newcommand{\rB}{\ensuremath{r_\B}\xspace}

\newcommand{\rBDK}{\ensuremath{r_\B^{\D \kaon}}\xspace}
\newcommand{\rBDpi}{\ensuremath{r_\B^{\D\pi}}\xspace}

\newcommand{\dB}{\ensuremath{\delta_\B}\xspace}
\newcommand{\dBDK}{\ensuremath{\delta_\B^{\D\kaon}}\xspace}
\newcommand{\dBDpi}{\ensuremath{\delta_\B^{\D\pi}}\xspace}

\renewcommand{\ci}{\ensuremath{c_i}\xspace}
\newcommand{\si}{\ensuremath{s_i}\xspace}

\newcommand{\Fi}{\ensuremath{F_i}\xspace}

\hypersetup{
  colorlinks   = true, 
  urlcolor     = blue, 
  linkcolor    = blue, 
  citecolor    = red   
}

\ifEnableSectionTOCLinks
    \usepackage[explicit]{titlesec} 
    
    \let\oldcontentsline\contentsline
    \renewcommand

    \titleformat{\section}{\normalfont\Large\bf}{\hyperlink{tocsection.\thesection}{{\thesection} \parbox[t]{\dimexpr\textwidth-1pc}{#1}}}{1pc}{}

    \titleformat{\subsection}{\normalfont\bf}{\hyperlink{tocsubsection.\thesubsection}{{\thesubsection} \parbox[t]{\dimexpr\textwidth-1pc}{#1}}}{1pc}{}

    \titleformat{name=\section,numberless}[display]{}{}{0pt}{\normalfont\Huge\bfseries #1}
\fi

\usepackage{cite} 
\usepackage{LHCb/mciteplus}

\usepackage{longtable} 
\usepackage{booktabs}

\begin{document}

\renewcommand{\thefootnote}{\fnsymbol{footnote}}
\setcounter{footnote}{1}


\begin{titlepage}
\pagenumbering{roman}

\vspace*{-1.5cm}
\centerline{\large EUROPEAN ORGANIZATION FOR NUCLEAR RESEARCH (CERN)}
\vspace*{1.5cm}
\noindent
\begin{tabular*}{\linewidth}{lc@{\extracolsep{\fill}}r@{\extracolsep{0pt}}}
\ifthenelse{\boolean{pdflatex}}
{\vspace*{-1.5cm}\mbox{\!\!\!\includegraphics[width=.14\textwidth]{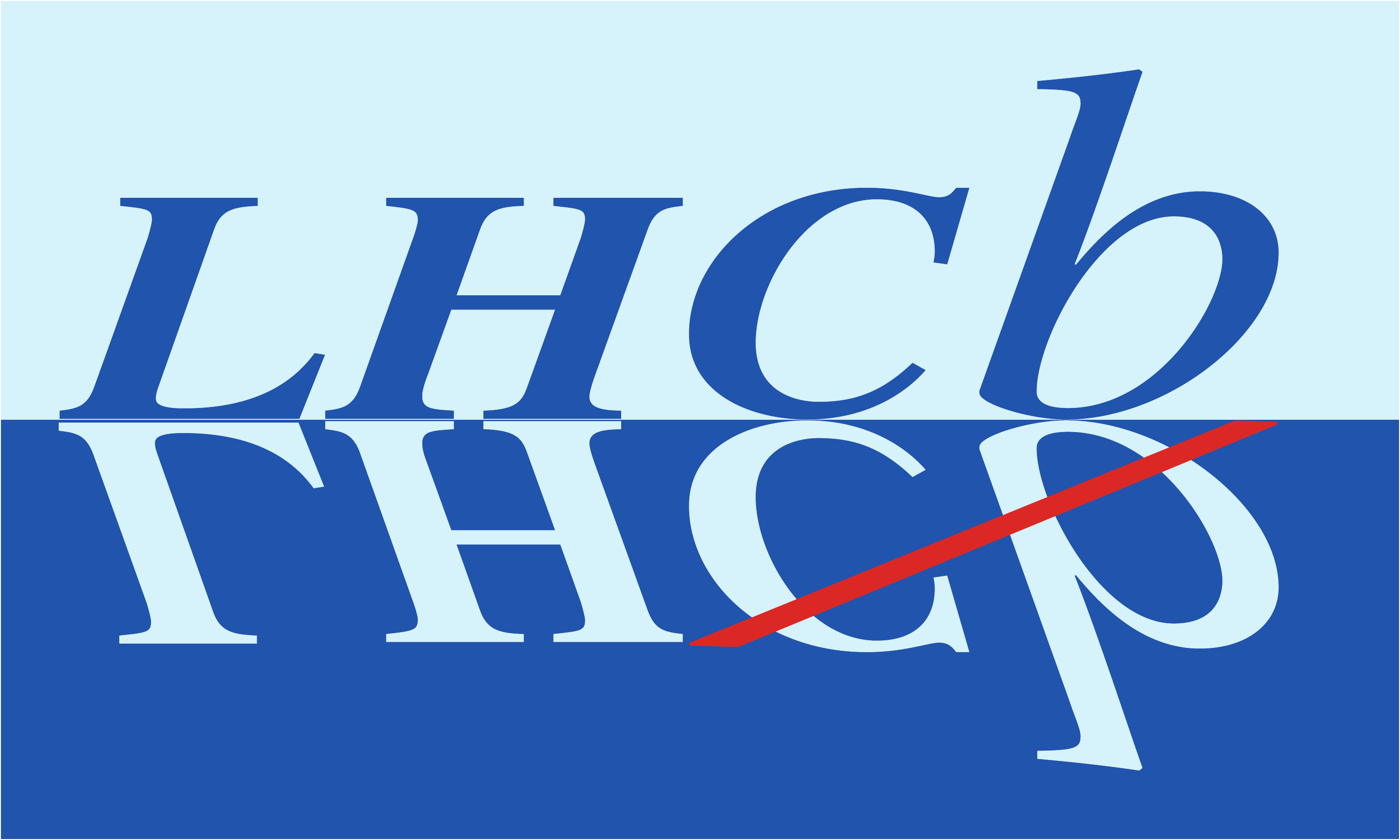}} & &}%
{\vspace*{-1.2cm}\mbox{\!\!\!\includegraphics[width=.12\textwidth]{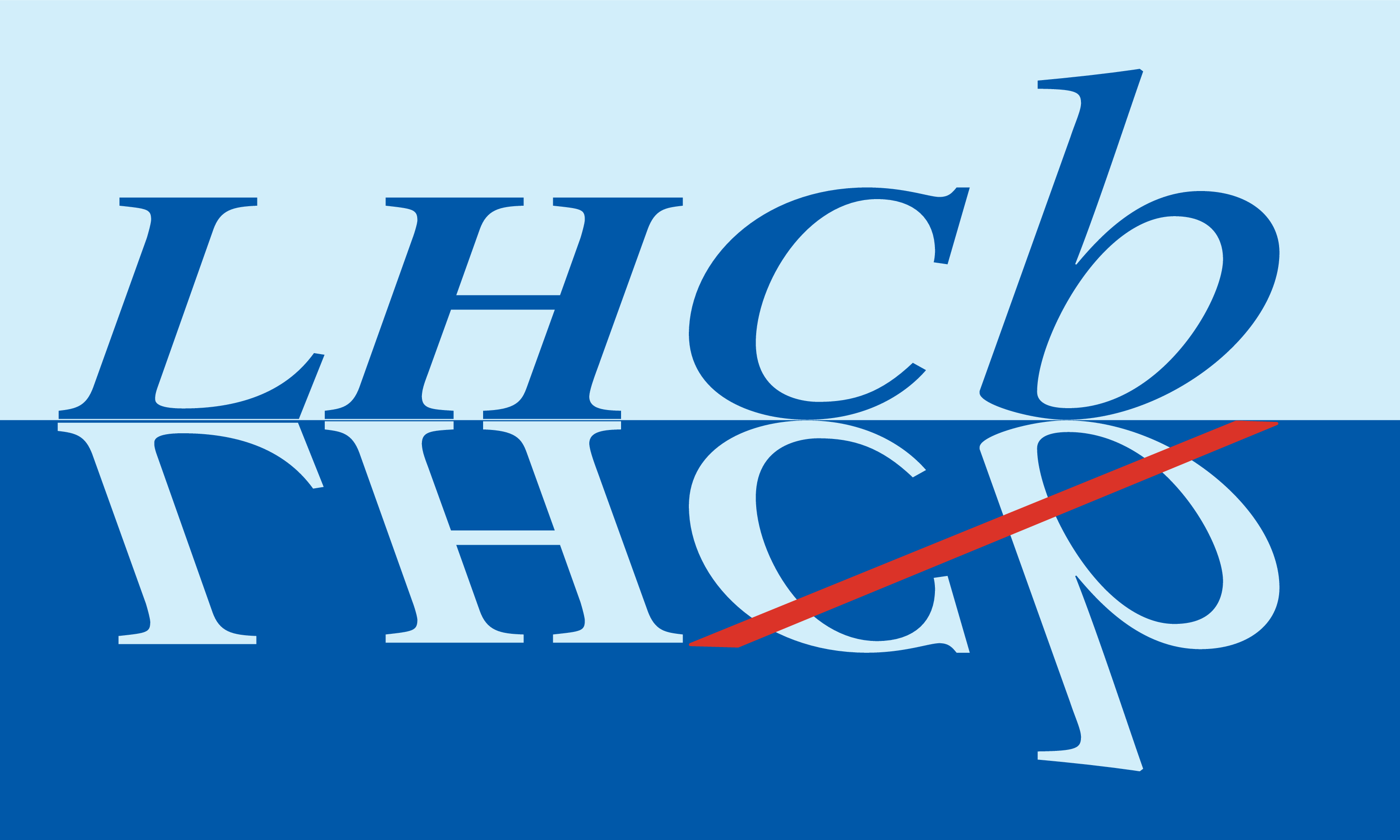}} & &}%
\\
 & & CERN-EP-2026-110 \\  
 & & LHCb-PAPER-2026-010 \\  
 & & August 10, 2026 \\ 
 & & \\
\end{tabular*}

\vspace*{4.0cm}

{\normalfont\bfseries\boldmath\huge
\begin{center}
  \papertitle 
\end{center}
}

\vspace*{2.0cm}

\begin{center}
\paperauthors\footnote{Authors are listed at the end of this paper.}
\end{center}

\vspace{\fill}

\begin{abstract}
  \noindent
    A measurement of the CKM angle \g using the decay channels \BtoDK and \BtoDpi, where the \D meson decays to \DtoKspp or \DtoKskk, is performed with a data sample corresponding to an integrated luminosity of 5.8\invfb, collected during 2024 by the upgraded \lhcb experiment. \CP violation is observed through a difference in the distributions of the Dalitz plot of the \D decay between the \Bp and \Bm mesons.  The CKM angle \g is determined to be $\g=(68.1\pm 6.7)^\circ$. Other parameters related to the examined \B meson decay modes are also measured. This is the first measurement of the CKM angle \g using the upgraded LHCb detector.
  
\end{abstract}

\vspace*{2.0cm}

\begin{center}
  Published in JHEP 08 (2026) 037
\end{center}

\vspace{\fill}

{\footnotesize 
\centerline{\copyright~\papercopyright. \href{\paperlicenceurl}{\paperlicence}.}}
\vspace*{2mm}

\end{titlepage}


\newpage
\setcounter{page}{2}
\mbox{~}


\renewcommand{\thefootnote}{\arabic{footnote}}
\setcounter{footnote}{0}


\cleardoublepage


\pagestyle{plain} 
\setcounter{page}{1}
\pagenumbering{arabic}

\section{Introduction}
\label{sec:Introduction}

The origin of \CP violation in the quark sector is described in the Standard Model (SM) through the elements of the CKM matrix~\cite{Cabibbo:1963yz,Kobayashi:1973fv}. A long-standing goal of particle physics is to overconstrain the lengths and angles of the Unitary Triangle, which is constructed from a subset of the CKM matrix elements and provides a geometric representation of quark mixing and \CP violation. Experimental evidence of nonclosure of the triangle would signify physics beyond the SM. The angle of the Unitary Triangle that is denoted as $\g \equiv \textrm{arg}\left(-\Vud\Vubs/\Vcd\Vcbs\right)$, which is the only CKM angle that can be measured in decays without any significant loop contribution,  is experimentally accessible through the interference of $\bquarkbar \to \cquarkbar\uquark\squarkbar$ and $\bquarkbar \to \uquarkbar\cquark\squarkbar$ (and \CP-conjugate) decay amplitudes. In addition, there are negligible theoretical uncertainties~\cite{BrodZup} when interpreting the measured observables in terms of the SM, as the parameters related to the strong interaction between the quarks in the decay can be measured alongside \g, with no need for external QCD input. Hence, in the absence of unknown physics effects at tree level, a precision measurement of \g provides a SM benchmark that can be compared with indirect determinations from other CKM-matrix observables more likely to be affected by physics beyond the SM~\cite{Blanke:2018cya}. 

In the last decade, the precision of direct determinations of  \g has been driven by the \lhcb experiment. A wide range of channels have been analysed and the combined direct determination from LHCb measurements is $(62.8\pm 2.6)^\circ$~\cite{LHCb-CONF-2025-003}, which is consistent with indirect determinations of \g from fits to the other CKM-matrix observables such as $(66.29^{+0.72}_{-1.86})^{\circ}$~\cite{CKMfitter2005,CKMfitter2015} and $(64.9\pm 1.4)^{\circ}$~\cite{UTfit-UT,UTfit:2022hsi}. 

The effects of interference between $\bquarkbar \to \uquarkbar\cquark\squarkbar$ and $\bquarkbar \to \cquarkbar\uquark\squarkbar$ (and \CP-conjugate) decay amplitudes can be observed in the \BtoDK decay, where the symbol \D is used to represent a superposition of \Dz and \Dzb states that decay to the same final state. In the case of $\D\to \Kspipi$,  \CP violation due to the phase $\gamma$ is observable in the Dalitz-plot distribution of the \D decay. Taking into account the \CP transformation, the distributions differ for signal events originating from \Bp and \Bm mesons. The interpretation of these differences within the context of the SM is discussed in further detail in Sect.~\ref{sec:theory}, drawing from the discussion in Refs.~\cite{BONDARGGSZ, GGSZ, AGGSZ}. The most precise measurements of $\gamma$ to date have extended this technique by including the \BtoDpi and \DtoKskk decays~\cite{LHCb-PAPER-2020-019,LHCb-PAPER-2025-063,LHCb-PAPER-2025-064}. 

This article presents a measurement using a similar model-independent method to that presented in Ref.~\cite{LHCb-PAPER-2020-019} but uses data accumulated by the \lhcb Run 3 detector during 2024 in $pp$ collisions at a centre-of-mass energy of $\sqs = 13.6 \tev$. The data correspond to a total integrated luminosity of approximately 5.8\invfb and are independent of the data used in previous measurements.

The analysis approach is laid out in Sect.~\ref{sec:theory}, while Sect.~\ref{sec:detector} describes the \lhcb Run 3 detector used to collect the data sample, and Sect.~\ref{sec:selection} summarises the  selection criteria. The measurement is based on a two-step fit procedure covered in Sect.~\ref{sec:massfit}, where the fit to the $B$-candidate mass distributions is detailed, and Sect.~\ref{sec:cpfit}, which describes how the \CP observables are determined through a simultaneous fit to the $B$-candidate mass in regions of the \D-decay phase space. The systematic uncertainties are reported in Sect.~\ref{sec:syst}, and the results are interpreted to determine the value of $\gamma$ in Sect.~\ref{sec:interpretation}. Finally, the conclusions are presented in Sect.~\ref{sec:conclusions}.

\section{Measurement Overview}

\label{sec:theory}

The analysis uses \BtoDK and \BtoDpi decays, where the \D meson decays to \Kspp and \Kskk. The analysis strategy is the same as that described in Ref.~\cite{LHCb-PAPER-2020-019}. The analysis measures six observables. Those related to the \BtoDK decay are
\begin{equation}
\xpmdk \equiv \rBDK \cos (\dBDK \pm \gamma) {\rm\ \ and\ } \; \ypmdk \equiv \rBDK \sin (\dBDK \pm \gamma),
\label{eq:xydefinitions}
\end{equation}
where the parameter \rBDK denotes the ratio of the suppressed $\Bm \to \Dzb \Km$ and the favoured $\Bm \to \Dz \Km$ decay-amplitude magnitudes (which is equal for the charge conjugate processes), and \dBDK denotes the relative \CP-conserving strong phase between the two amplitudes. The size of the interference is proportional to $\rBDK\approx0.097$~\cite{LHCb-CONF-2025-003}. The \BtoDpi and \BtoDK decays both share the same weak phase \g, but have different amplitude ratios and strong-phase differences. Instead of introducing four analogous parameters to account for these differences, \CP violation in \BtoDpi decays can be parametrised by a single additional complex variable~\cite{Tico:2019xdx}. This variable is
\begin{align}
    \xi^{\dpi}=\left(\frac{\rBDpi}{\rBDK}\right)\exp{\left(i\dBDpi-i\dBDK\right)},
    \label{eq:xidefinitions}
\end{align}
leading to two additional observables $x_\xi^{\dpi}\equiv\Re (\xi^{\dpi})$ and $y_\xi^{\dpi}\equiv\Im (\xi^{\dpi})$. The parameters \xpmdpi and \ypmdpi can be derived through the relations
\begin{align}\label{eq:xy_from_xi}
    x_\pm^{\D\pi} &= x_\xi^{\D\pi}\xpmdk - y_\xi^{\D\pi}\ypmdk, 
    & y_\pm^{\D\pi} &= x_\xi^{\D\pi}\ypmdk + y_\xi^{\D\pi}\xpmdk.
\end{align} 
The \BtoDpi decay has a branching fraction approximately 13 times larger than the \BtoDK decay. However, as  $\rBDpi \approx 0.005$ is much smaller than $\rBDK$~\cite{LHCb-CONF-2025-003},  the \BtoDpi channel is primarily used as a control mode, although its weak sensitivity to \g is incorporated into the results. 

\begin{figure}[t]
\begin{center}
\includegraphics[width=0.48\textwidth]{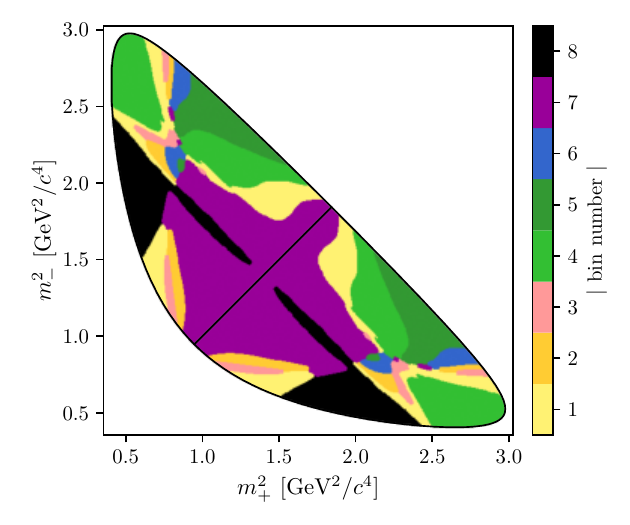}
\includegraphics[width=0.48\textwidth]{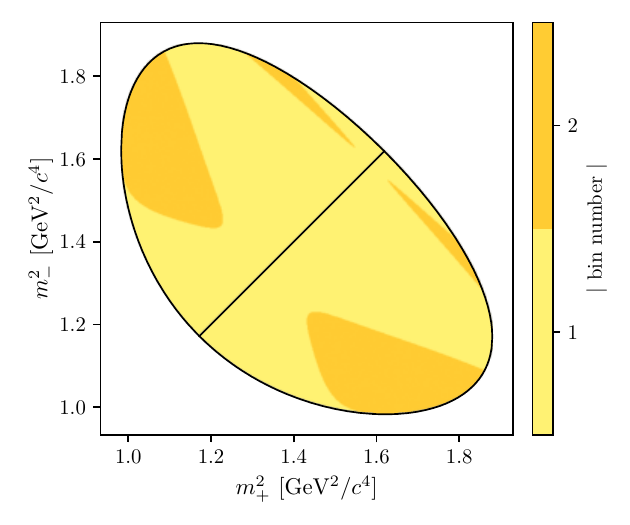}
\caption{\small Binning schemes for (left) \DtoKspp decays and (right) \DtoKskk decays. The diagonal line separates the positive and negative bins, where the positive bins are in the region in which $\msqmin > \msqplus$ is satisfied. }
\label{fig:bins}
\end{center}
\end{figure}

The analysis proceeds through the examination of the signal event distribution over the $\D$-decay phase space. The position on the Dalitz plot is defined by the coordinates $\msqmin$ and $\msqplus$, which are the squared masses of the $\KS h^-$ and  $\KS h^+$ particle combinations, where $h$ is a pion or a kaon. The $\Dz \to \KS h^+ h^-$ decay amplitude is denoted by  $A_D(\msqmin,\msqplus)$ and that of $\Dzb \to \KS h^+ h^-$ by $A_{\Dbar}(\msqmin,\msqplus)$. The analysis partitions the $D$-decay phase space into regions that are symmetric around  $\msqmin=\msqplus$, resulting in $2 \times \mathcal{N}$ bins labelled from $i= -\mathcal{N}$ to $i= +\mathcal{N}$ (excluding zero). Bins for which $\msqmin > \msqplus$ are defined to have positive values of $i$. The binning schemes used in this analysis are depicted in Fig.~\ref{fig:bins} and are defined in Ref.~\cite{CLEOCISI}. For \DtoKspp the scheme named \emph{optimal} is used, whereas for \DtoKskk the scheme that is used is named \emph{2 bins}. These schemes are determined from amplitude models~\cite{CLEOCISI} and are chosen from those available in order to maximise the expected sensitivity to \g, given anticipated yield and purity at the \lhcb experiment, based on knowledge from Run~1 and 2 analyses.

Selection requirements and reconstruction effects lead to a nonuniform efficiency over the phase space, denoted by $\epsilon(\msqmin, \msqplus)$. In order to reduce the reliance of the analysis on simulation, the effect of $\epsilon(\msqmin, \msqplus)$ is incorporated into the analysis rather than corrected for. The fractional yield of pure \Dz decays in bin $i$ in the presence of this efficiency profile is denoted as
\begin{equation}
\label{eq:fi}
\Fi = \frac{\int_{i} \deriv\msqmin\, \deriv\msqplus\, |A_{D}(\msqmin,\msqplus)|^2 \, \epsilon(\msqmin,\msqplus) }{\sum_j \int_{j} \deriv\msqmin\, \deriv\msqplus\, |A_{D}(\msqmin,\msqplus)|^2\, \epsilon(\msqmin,\msqplus) },
\end{equation}
where the sum in the denominator is over all Dalitz-plot bins, indexed by $j$. The analysis ignores \CP violation in charm decays, the effects of charm mixing, as well as the presence of \CP violation and matter regeneration in the neutral $K^0$ decay. These effects are expected to have a small impact~\cite{BPV,KsCPV} on the distribution of events over the Dalitz plot, and the impact of these assumptions is accounted for as a systematic uncertainty. With these assumptions, the charge-conjugate amplitudes satisfy the relation $A_{\Dbar}(\msqmin,\msqplus) = A_D(\msqplus,\msqmin)$, and therefore $F_{+i}$ = $\overline{F}_{-i}$, where $\overline{F}_i$ is the fractional yield of observed \Dzb decays in bin $i$. It is assumed that the relative change in $\epsilon(\msqmin, \msqplus)$ over the Dalitz plot is the same for both \BtoDpi and \BtoDK decays. This assumption allows the same $\Fi$ to be used for both $B$ decay channels and is verified using simulation.

The amplitude of the $\Bm \to\D\Km$ decay is the sum of the favoured (via \Dz) and suppressed (via \Dzb) contributions and is given by
\begin{equation}
A_B (\msqmin,\msqplus) \propto \, A_D(\msqmin,\msqplus) + \rBDK e^{i(\dBDK - \g)}A_{\Dbar}(\msqmin,\msqplus).
\label{eq:bamplitude}
\end{equation}
The equivalent expression for the charge-conjugated decay $B^+ \to D K^+$ is obtained by making the substitutions $\gamma \leftrightarrow -\gamma$ and $A_D(\msqmin,\msqplus) \leftrightarrow A_{\Dbar}(\msqmin,\msqplus)$.

The strong-phase difference between the \Dz- and \Dzb-decay amplitudes at a given point on the Dalitz plot is denoted as $\delta_D(m_-^2,m_+^2)$.
The cosine of $\delta_D(m_-^2,m_+^2)$ weighted by the \D-decay amplitude and averaged
over bin $i$ is denoted by $c_i$~\cite{GGSZ}, and is given by
\begin{align}
c_i &\equiv \frac{\int_{i} \deriv\msqmin \, \deriv\msqplus \, |A_D(\msqmin,\msqplus)| |A_D(\msqplus,\msqmin)| \cos\
[\delta_D(\msqmin,\msqplus)]}
{\sqrt{\int_{i} \deriv\msqmin \, \deriv\msqplus \, |A_D(\msqmin,\msqplus)|^2 \int_{i} \deriv\msqmin \, \deriv\msqplus \, |A_D(\msqplus,\msqmin)|^2}}\,,
\label{eq:ci}
\end{align}
where the integrals are evaluated over the phase-space region defined by bin $i$. An analogous expression can be written for $s_i$, which is the sine of the strong-phase difference weighted by the decay amplitude and averaged over the bin phase space. In Eq.~\eqref{eq:ci} the impact of nonuniform efficiency is ignored, because the only measurements available are those of \ci and \si\cite{CLEOCISI,bes3prd,bes3prl,BESIII:2025nsp,Krishna}, and not those modulated by the \lhcb efficiency profile. The impact of this simplification is investigated as a source of systematic uncertainty.

The expected yield of \Bm decays in bin $i$ is found by integrating the square of the amplitude given in  Eq.~\eqref{eq:bamplitude} over the region of phase space defined by the $i$-th bin.  
Using the relations $c_{i}\equiv c_{+i} = c_{-i}$ and $s_{i}\equiv s_{+i}= - s_{-i}$, the $\Bp \to \D K^+$ ($\Bm \to \D K^-$) yields, $N^+$ ($N^-$), in bins $+i$ and $-i$ are given by
\begin{align}
\begin{split}
N_{+i}^+ &= h_{B^+}^{\D K} \left[ F_{-i} + \left(\left(\xpdk\right)^2 + \left(\ypdk\right)^2\right) F_{+i} + 2 \sqrt{F_{+i} F_{-i}} \left( \xpdk c_{+i} - \ypdk s_{+i}\right) \right], \\
N_{-i}^+ &= h_{B^+}^{\D K} \left[ F_{+i} + \left(\left(\xpdk\right)^2 + \left(\ypdk\right)^2\right) F_{-i} + 2 \sqrt{F_{+i} F_{-i}} \left( \xpdk c_{+i} + \ypdk s_{+i}\right) \right], \\
N_{+i}^- &= h_{B^-}^{\D K} \left[ F_{+i} + \left(\left(\xmdk\right)^2 + \left(\ymdk\right)^2\right) F_{-i} + 2 \sqrt{F_{+i} F_{-i}} \left( \xmdk c_{+i} + \ymdk s_{+i}\right) \right], \\
N_{-i}^- &= h_{B^-}^{\D K} \left[ F_{-i} + \left(\left(\xmdk\right)^2 + \left(\ymdk\right)^2\right) F_{+i} + 2 \sqrt{F_{+i} F_{-i}} \left( \xmdk c_{+i} - \ymdk s_{+i}\right) \right], 
\label{eq:populations}
\end{split}
\end{align}
where $h_{B^+}^{\D K}$ and $h_{B^-}^{\D K}$ are normalisation constants. The value of \rBDK is effectively independent for each $B$ meson charge as it can be constructed from either $\sqrt{\xp^2+\yp^2}$ or $\sqrt{\xm^2+\ym^2}.$ This choice reduces correlation among the \CP observables. When the \CP observables are subsequently interpreted to determine the physics parameters of interest, only a single value of $\rBDK$ is determined. Analogous equations for the \BtoDpi decay yields can be written with the substitutions $\xpmdk \rightarrow \xpmdpi$ and $\ypmdk \rightarrow \ypmdpi$ and the introduction of two further normalisation constants $h_{B^+}^{\D \pi}$ and $h_{B^-}^{\D \pi}$. For the \DtoKshh decay channels, any phase-space-integrated asymmetry due to \g is expected to be small and heavily biased due to the effects of \KS ~\CP violation~\cite{KsCPV}. Therefore the use of four independent normalisation constants purposefully ignores these asymmetries. This is advantageous as in this format the effects of the production asymmetry of \Bpm mesons in $pp$ collisions and the detection asymmetries of the charged companion from the \B decay are absorbed into the normalisation constants. This leads to a \CP-violation measurement that is free of systematic uncertainties associated with these effects. 

Across the system of analysed decay channels, the observed yields, the $N^+$ and $N^-$ in each bin for both the \BtoDK and \BtoDpi decays, outnumber the unknown parameters (the \CP parameters, the normalisation constants, the \ci, \si, and \Fi). Nonetheless, 
external information on the strong-phase parameters \ci and \si is used to enhance the precision on the \CP observables. Measurements of \ci and \si are best made using pairs of neutral charm mesons that are quantum correlated, which allow direct determinations of the strong-phase difference between the \Dz and \Dzb decays. Measurements have been made by the \besiii collaboration using techniques developed at \cleo. For the \DtoKspp decay, the values of \ci and \si are taken from Ref.~\cite{BESIII:2025nsp} which uses 7.93\invfb of data collected at \besiii. Multiple results are presented in Ref.~\cite{BESIII:2025nsp}; the ones used here are those which do not impose weak-model constraints between the $\Dz \to\KS\pip\pim$ and $\Dz \to \KL \pip\pim$ decay amplitudes, in contrast to previous measurements~\cite{bes3prd,bes3prl} which imposed constraints based on the assumption of U-spin symmetry. For the \DtoKskk events, the \ci and \si measurements are from the combination of \besiii~\cite{Krishna} and \cleo~\cite{CLEOCISI} data detailed in  Ref.~\cite{Krishna}. A significant benefit of using direct measurements rather than amplitude-model derived values of \ci and \si is that propagation of measurement uncertainty is relatively straightforward to perform, in contrast to estimation of systematic uncertainties on the phase from models. Due to a phase-convention difference between the \lhcb and \besiii experiments it is necessary to multiply the \si values in~\cite{BESIII:2025nsp,Krishna} by ($-1$) for use in this analysis.

The extraction of the \xpmdk, \ypmdk, \xxidpi, \yxidpi and \Fi parameters is performed using a simultaneous fit to the  \BtoDpi and \BtoDK candidate invariant-mass spectra in each Dalitz-plot bin. Due to the relative values of \rBDpi and \rBDK and the relative branching fractions of the two \B decay modes, it is expected that the \xpmdk and \ypmdk observables are primarily determined by the \BtoDK decay whereas the \Fi parameters are primarily determined by the \BtoDpi decay.

\section{LHCb Detector}

\label{sec:detector}

The \lhcb Run 3 detector~\cite{LHCb-DP-2022-002} is a single-arm forward spectrometer covering the pseudo\-rapidity range $2<\eta <5$, designed for 
the study of particles containing \bquark or \cquark quarks. 
It was installed during LHC long-shutdown 2, to allow effective operation at about five times the luminosity with respect to the Run 1--2 detector~\cite{LHCb-DP-2022-002,LHCb-DP-2008-001}.
The LHCb Run~3 detector keeps the general organisation of subsystems of the original LHCb detector although most subsystems were changed.

The high-precision tracking system consists 
of a silicon-pixel vertex detector surrounding the $pp$
interaction region, a large-area silicon-strip detector located upstream of a dipole magnet with a bending power of about $4{\mathrm{\,T\,m}}$, and three stations of scintillating-fibre detectors.
The magnet polarity is flipped on a regular basis, and 72\% (28\%) of the data in this analysis are collected under the configuration with the magnetic field pointing upwards (downwards).
Different types of charged hadrons are distinguished using information
from two ring-imaging Cherenkov detectors, equipped with  photon detection systems.
Photons, electrons and hadrons are identified by a calorimeter system consisting of electromagnetic
and hadronic calorimeters. Muons are identified by a
system composed of alternating layers of iron and multiwire
proportional chambers.

The readout of all LHCb subsystems at the bunch-crossing frequency and the subsequent processing of the data in an all-software trigger~\cite{LHCb-TDR-016, LHCb-TDR-017, LHCb-TDR-018, LHCb-TDR-021} is a central feature of the upgraded detector, enabling the reconstruction and selection of events in real time.  
The trigger system is implemented in two stages: a GPU-based inclusive stage (HLT1) focused primarily on charged particle reconstruction, which reduces the data volume by roughly a factor of 20~\cite{LHCb-TDR-021}, followed by a CPU-based stage (HLT2), which performs the physics-analysis-quality reconstruction and selection of physics signatures~\cite{LHCb-TDR-016}. A large disk buffer is placed between these stages to hold the data while the real-time alignment and calibration is being performed.
Owing to the removal of the hardware trigger present during Run~1 and Run~2 and the new optimised inclusive HLT1 trigger, 
a higher reconstruction efficiency of low transverse momentum candidates is achieved for fully hadronic decays.
At HLT1, events are required to have either a single high-transverse-momentum track which is inconsistent with originating from the primary vertex (PV), 
or have a pair of tracks forming a good-quality displaced secondary vertex selected by a multivariate classifier~\cite{Kitouni:2021fkh}.
The measurement presented in this article also benefits from a series of HLT2 triggers that exclusively select the final state particles based on their topology. 
Triggered data further undergo a centralised, offline processing step
to deliver physics-analysis-ready data across the entire \lhcb physics programme~\cite{Sprucing, FunTuple}.

Simulation is required to optimise the selection criteria, model the $B$-candidate mass shapes and evaluate relative efficiencies among background and signal decays.
It is also used to validate several assumptions in the measurement and study the associated systematic uncertainties.
 In the simulation, $pp$ collisions are generated using \pythia~\cite{Sjostrand:2007gs,*Sjostrand:2006za} with a specific \lhcb configuration~\cite{LHCb-PROC-2010-056}.
Decays of unstable particles are described by \evtgen~\cite{Lange:2001uf}, in which final-state radiation is generated using \photos~\cite{davidson2015photos}.
The interaction of the generated particles with the detector, and its response, are implemented using the \geant toolkit~\cite{Allison:2006ve, *Agostinelli:2002hh} as described in Ref.~\cite{LHCb-PROC-2011-006}. Fast simulation~\cite{Cowan:2016tnm} is adopted when studying subdominant backgrounds that are ignored in the baseline model.

\section{Reconstruction and selection}
\label{sec:selection}

The data used in this analysis are a subset of those collected during 2024 operation of LHCb, corresponding to the data-taking period in which the mass resolution and alignment conditions were sufficiently consistent for this measurement.

Reconstruction of the signal decay channels starts with the identification of \KS candidates. Decays of \decay{\KS}{\pip\pim} are reconstructed in two different categories:
the first involving \KS mesons that decay early enough for the
pions to be reconstructed in the vertex detector; and the
second containing \KS that decay later such that track segments of the
pions cannot be formed in the vertex detector. These categories are
referred to as \emph{long} and \emph{downstream}, respectively. The
\emph{long} category has better mass, momentum and vertex resolution than the
\emph{downstream} category. The \KS candidates are combined with two oppositely charged tracks to form \D meson candidates. A \B candidate is then formed by combining the \D meson with another track. At each stage of the combination, selection requirements are placed to ensure good quality vertices, and the \KS and \D candidate invariant masses are required to be close to their known masses~\cite{PDG2024}. 
The invariant mass is kinematically constrained through a fit imposed on the full \Bpm decay chain~\cite{Hulsbergen:2005pu}, where the \D and \KS candidates are constrained to their known masses~\cite{PDG2020} and the \Bpm candidate momentum vector is required to point towards the associated primary vertex, which is chosen to be that minimising the $B$-candidate impact parameter. Candidates are split into mutually exclusive \BtoDpi and \BtoDK categories based on the value of a particle identification (PID) variable of the companion track from the $B$ meson which ensures no overlap between the categories. The value is chosen through an optimisation of the full analysis chain to minimise the expected uncertainty on \g from pseudoexperiments.
Candidates with a constrained-fit mass in the range [5080, 5800]\mevcc are retained for further study.

Misidentified backgrounds from semileptonic \D decays or other \DtoKshh decays are suppressed through PID requirements on the tracks. For the \emph{long} category the \KS decay vertex is required to be separated in the $z$-direction from the \D meson vertex by more than five standard deviations in order to remove background from $\D \to \pi^+\pi^- h^+ h^-$ decays. This requirement is exceeded by default in the \emph{downstream} category.  Companion tracks that are associated with hits in the muon detector are removed. Decays of $B$ mesons to the same final state as the signal channel can proceed without the intermediate \D meson. To suppress these backgrounds, the $D$ meson decay vertex is required to be separated in $z$ from the $B$ meson vertex by at least one standard deviation, where $z$ is the direction of the beam axis. Background events of this type are rare and therefore a more stringent requirement is not required. 

Combinatorial background is suppressed through the use of a boosted decision tree (BDT) multivariate classifier~\cite{Breiman, AdaBoost}. The training sample used as a proxy for the signal is taken from simulated events. The background training sample is formed from \BtoDpi candidates in the invariant-mass range [5450, 5800]\mevcc, away from the $\Bpm$ signal peak. To avoid overlap between training samples and candidates entering the fit, a $k$-fold method~\cite{kFold} with $k=2$, is used on the background sample. Variables that enter the training relate to the kinematics and topology of the signal decay. These include the momentum and transverse momentum of the \B, \D, and companion mesons, the radial distance of the \B and \D decay vertices from the $z$-axis, the quality of the \B, \D and \KS meson vertices, the maximum distance of closest approach between \B meson decay products and \D meson decay products, and resolution-weighted separation of the \B candidate vertex from the primary vertex. The final set of inputs are based on the \chisqip of the tracks, and the \D and \B mesons, where \chisqip\ is defined as 
the difference in the vertex-fit \chisq of a given PV reconstructed with and
without the particle under consideration.
Independent BDTs are trained for the \emph{long} and \emph{downstream} categories. The values for the BDT requirement are determined by optimising the sensitivity to \g in pseudoexperiments after optimising requirements on the PID variable. 

\section{The \boldmath{$\DK$} and \boldmath{$\Dpi$} invariant-mass spectra}
\label{sec:massfit}

\begin{figure}[htb]
    \centering
    \includegraphics[width=0.49\linewidth]{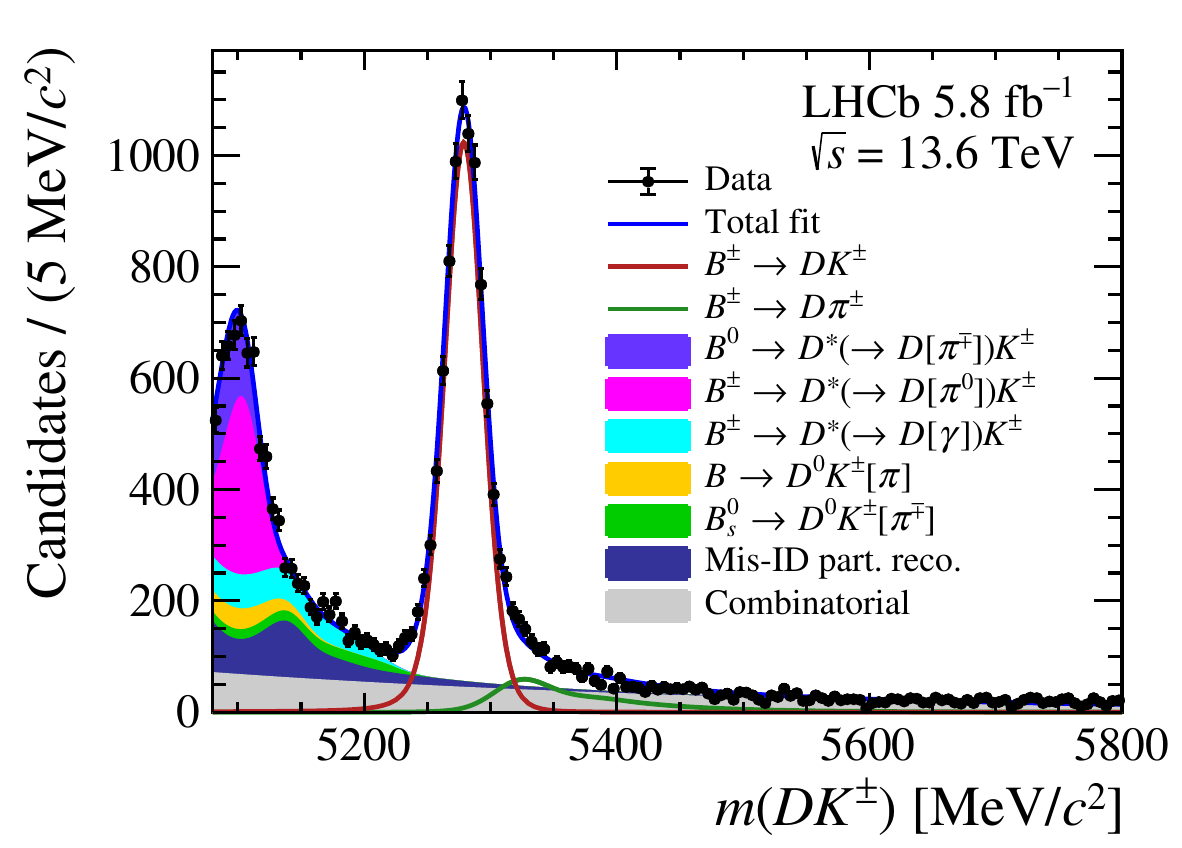}
    \includegraphics[width=0.49\linewidth]{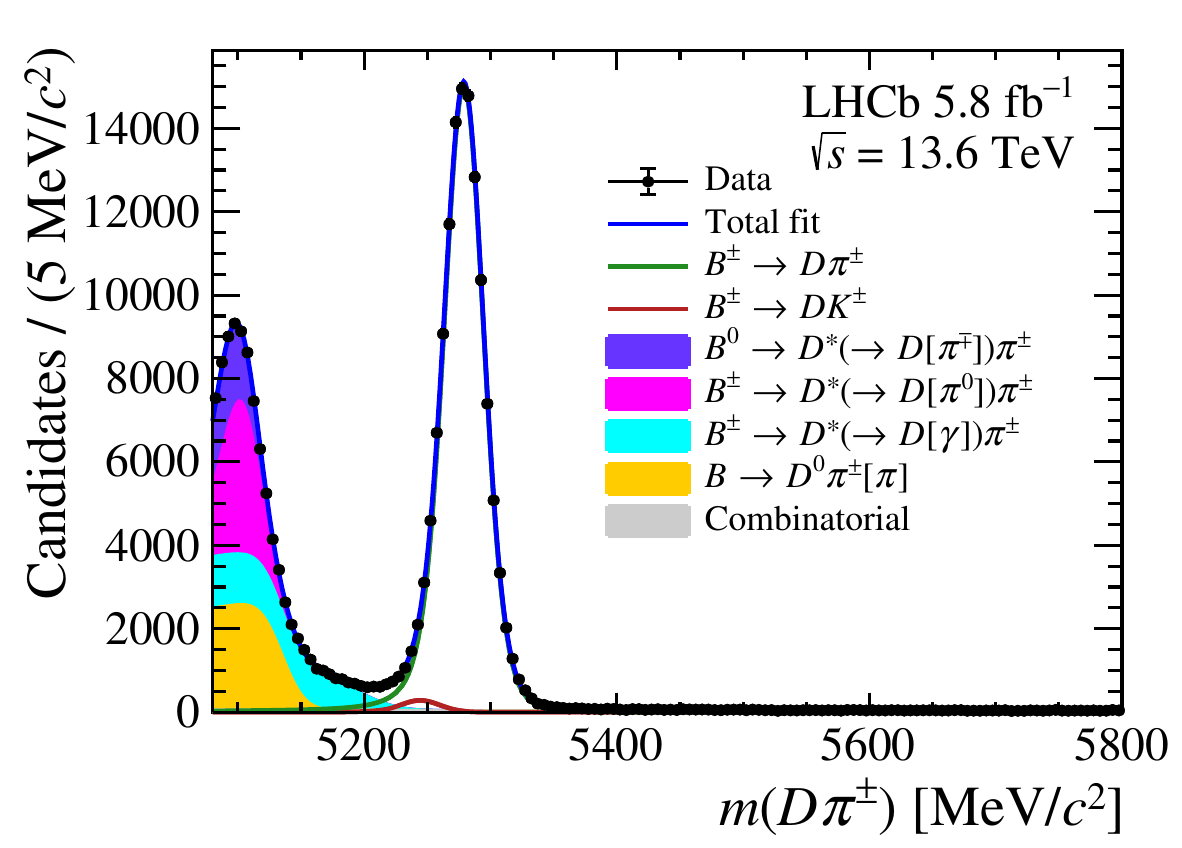}\\
    \includegraphics[width=0.49\linewidth]{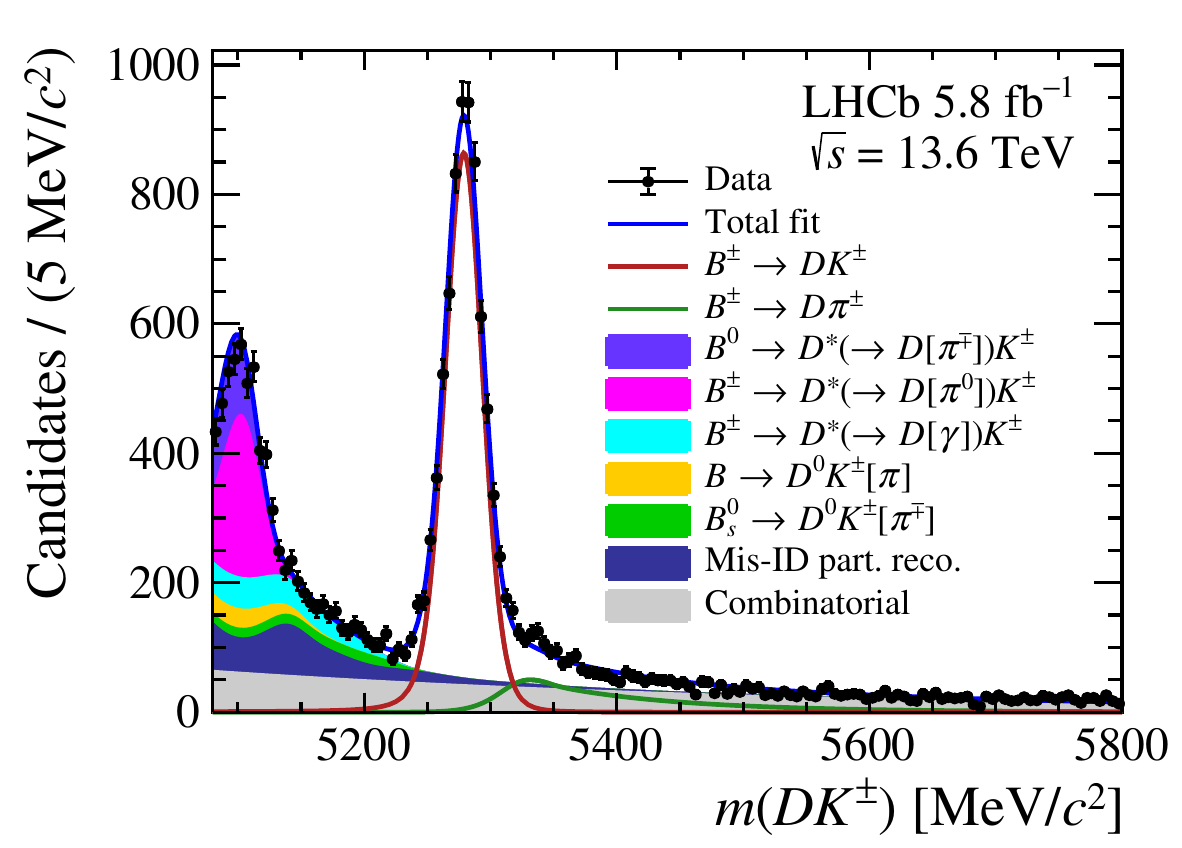}
    \includegraphics[width=0.49\linewidth]{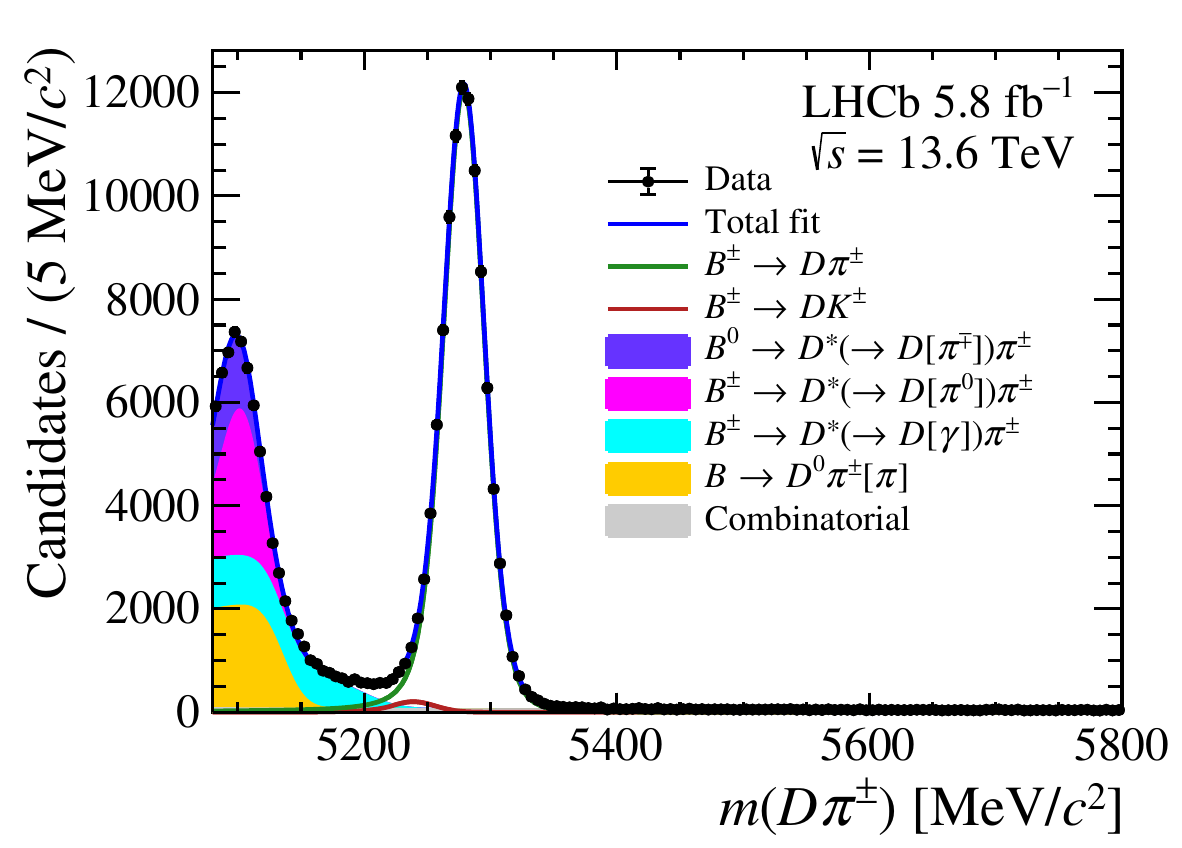}\\

    \caption{Distributions of the $Dh^{\pm}$ invariant mass for (left) $\BtoDK$ and (right) $\BtoDpi$ candidates, where $\DtoKspp$. The fit results are overlaid.
    The top row corresponds to the \emph{long} \KS reconstruction and the bottom row to the \emph{downstream} \KS reconstruction.}
    \label{fig:global_fit_res1}
\end{figure}

\begin{figure}[htb]
    \centering
    \includegraphics[width=0.49\linewidth]{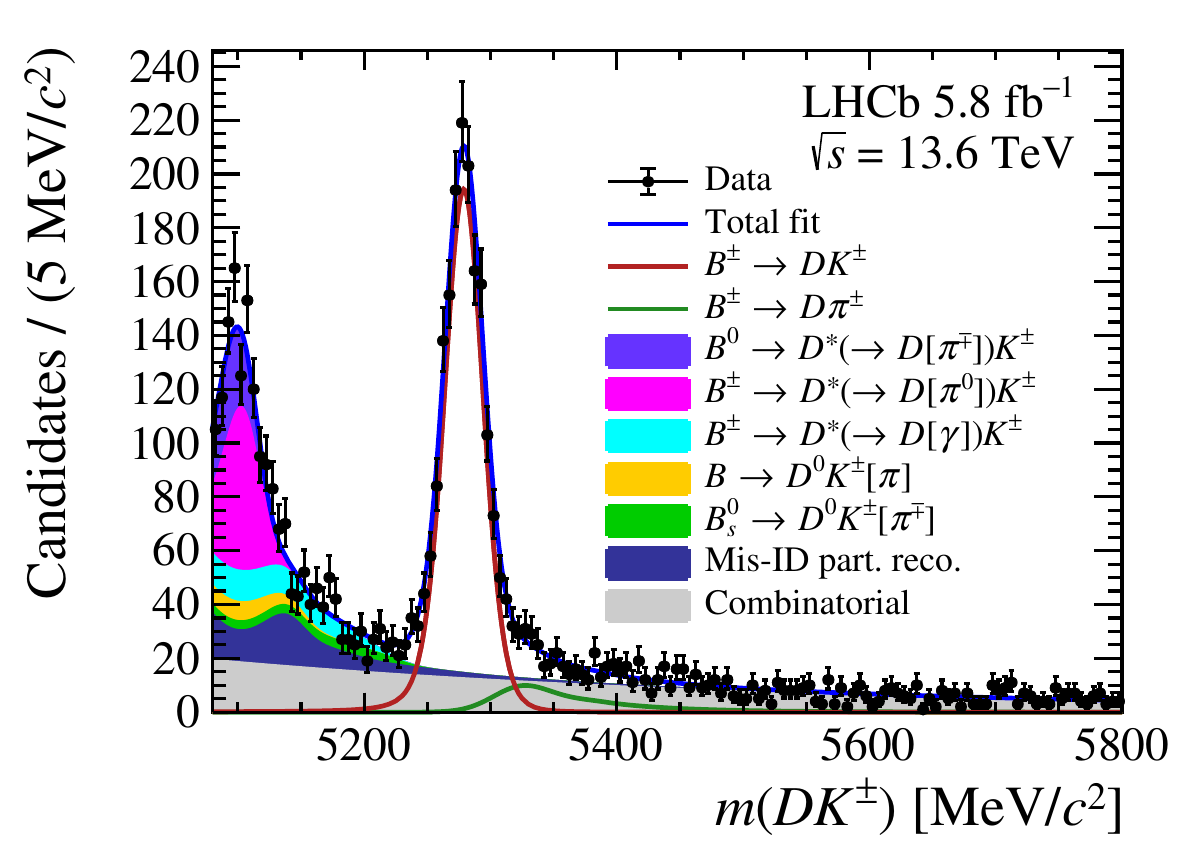}
    \includegraphics[width=0.49\linewidth]{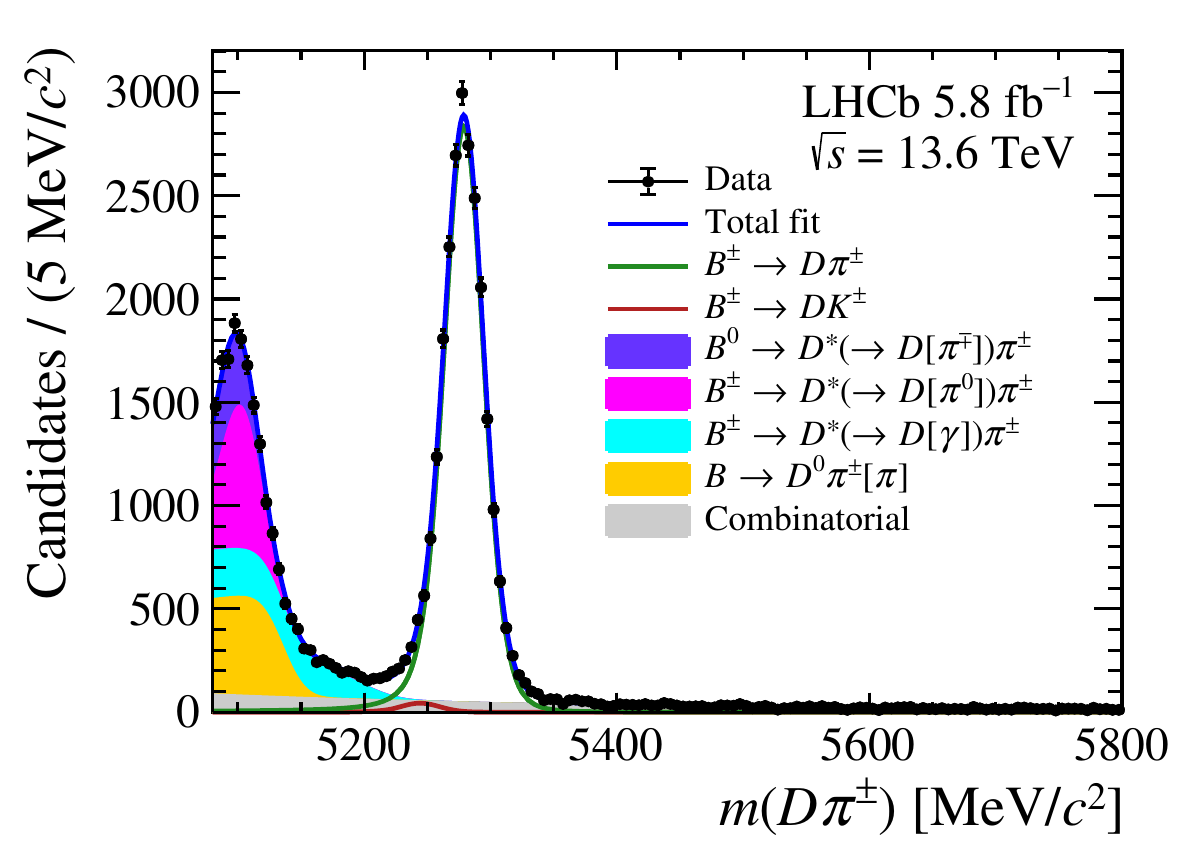}\\
    \includegraphics[width=0.49\linewidth]{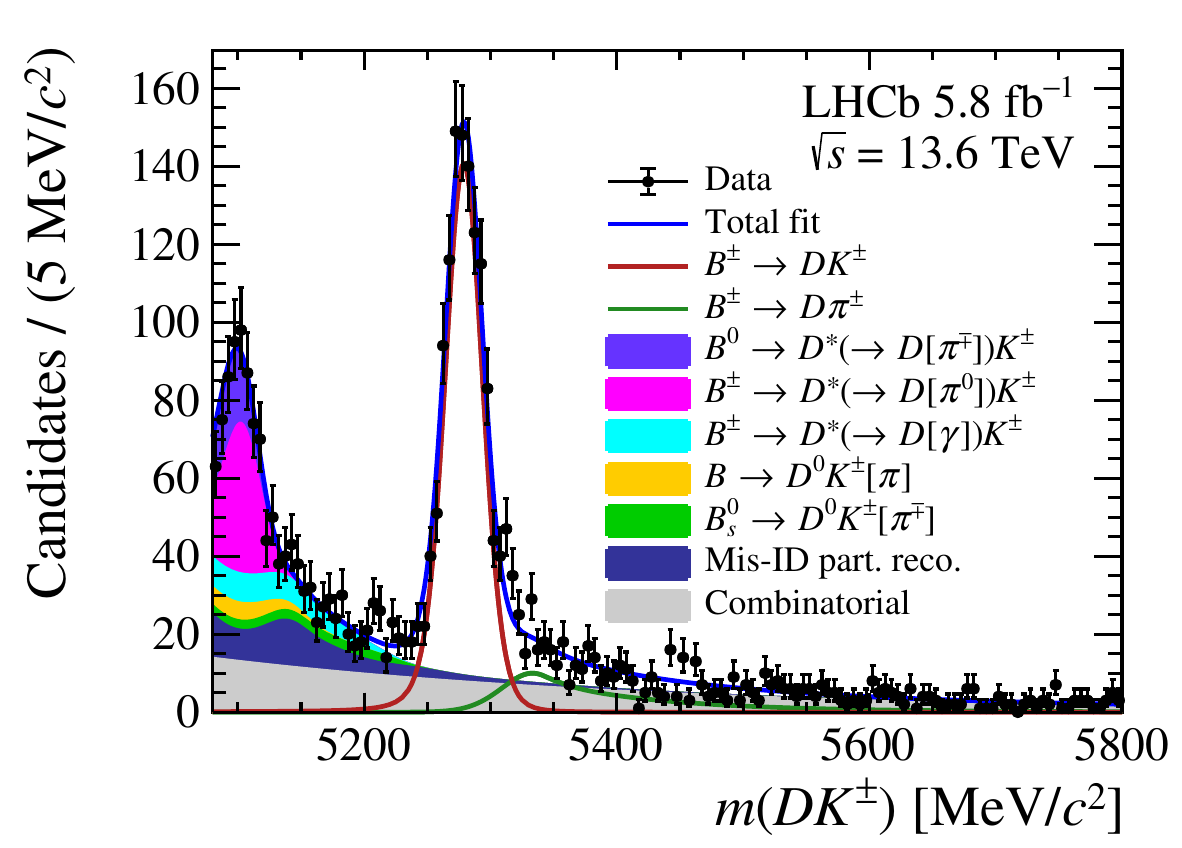}
    \includegraphics[width=0.49\linewidth]{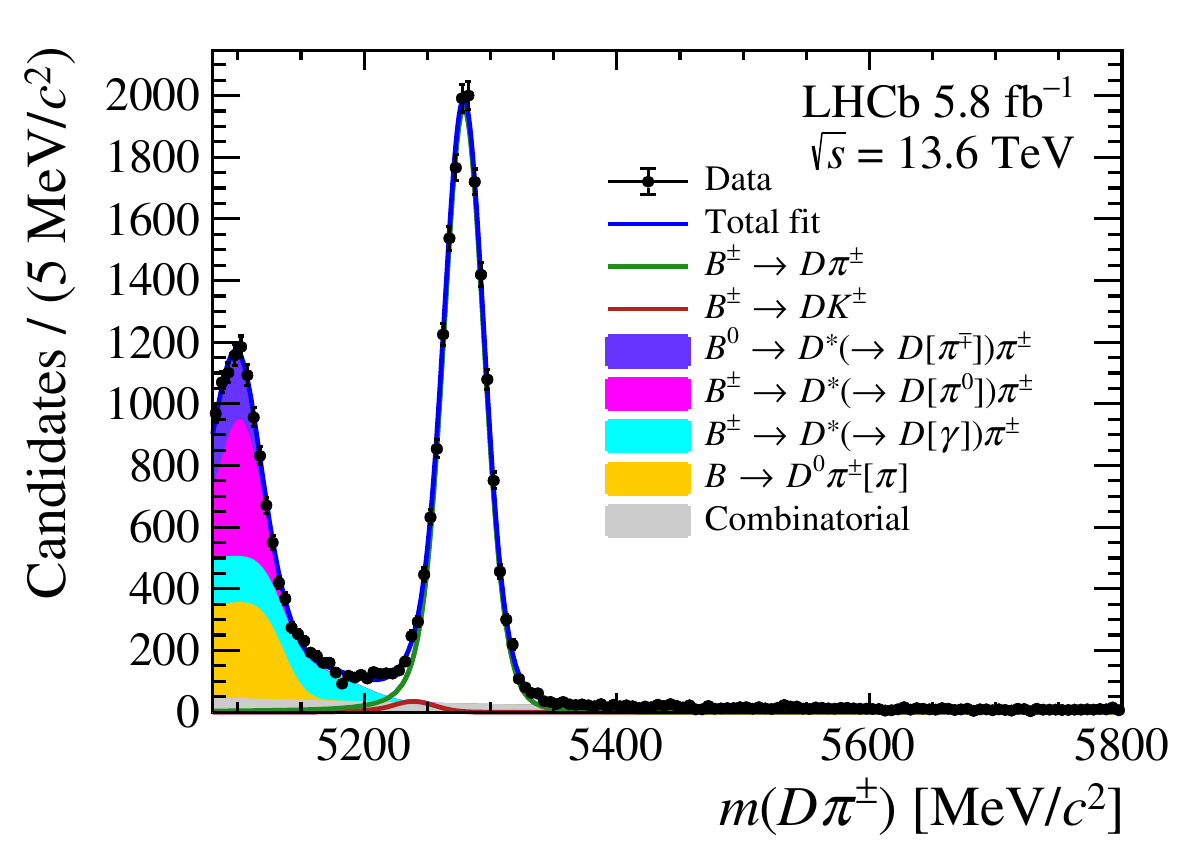}\\

    \caption{Distributions of the $Dh^{\pm}$ invariant mass for (left) $\BtoDK$ and (right) $\BtoDpi$ candidates, where $\DtoKskk$. The fit results are overlaid.
    The top row corresponds to the \emph{long} \KS reconstruction and the bottom row to the \emph{downstream} reconstruction.}
    \label{fig:global_fit_res2}
\end{figure}

The \CP observables are determined through a two-stage fit strategy. The first part is an unbinned extended maximum-likelihood fit to the invariant-mass spectra of all selected \Bpm candidates. This fit, referred to as the global mass fit, simultaneously determines the signal and background yields in the data across categories dependent on \B decay, \D decay, and \KS track type. The shape parametrisations determined in this fit are subsequently used in the second stage, where the data are further split by the charge of the \B meson and partitioned into the Dalitz-plot bins in order to determine the \CP observables. The invariant-mass distributions of selected \Bpm candidates are shown for \DtoKspp and \DtoKskk candidates in Figs.~\ref{fig:global_fit_res1} and \ref{fig:global_fit_res2}, together with the results of the fit superimposed.

The signal shape uses the same modified-Gaussian parametrisation as described in Ref.~\cite{LHCb-PAPER-2020-019}. The mass and width are allowed to vary in the fit to data with all other parameters fixed from simulation. The varying width accounts for the resolution differences between data and simulation. The peak position parameter is shared across all categories while the width is separate for each \B decay channel and \KS track type. The yield of the  \BtoDpi decay is determined independently in each \B decay, \D decay and \KS track type combination. The fit determines the ratio of the branching fractions for \BtoDK decays relative to \BtoDpi decays, $R_{DK/D\pi}$. The ratios of selection efficiencies determined from simulation in each category are included in the fit to unify this into a single fit parameter that is shared across \KS-decay categories and \D decays.

Candidates where the companion pion has been misidentified as a kaon in the \BtoDK category are primarily distributed to the right of the signal peak in the invariant-mass distribution. The rate of this decay is substantially reduced compared to the data in Ref.~\cite{LHCb-PAPER-2020-019} due to a tighter PID requirement following the optimisation described in Sect.~\ref{sec:selection}. The shape is determined from a fit to \BtoDpi candidates alone. The misidentified distribution of \BtoDpi candidates is found through application of the \sPlot~\cite{Pivk:2004ty} technique in order to subtract the small background contribution. The shape is then modified by applying the PID efficiency for each candidate as a function of the companion track momentum, pseudorapidity and number of long tracks in the event. The PID efficiency is determined from calibration data weighted to match the companion track in these variables. The probability density function (PDF) for this background is obtained by fitting the derived distribution with the sum of two Crystal Ball~\cite{Skwarnicki:1986xj} (CB) functions. All parameters are then fixed in the global fit. The complementary background where \BtoDK decays are misidentified as \BtoDpi decays is also present though at a smaller relative rate due to the lower branching fraction of \BtoDK decays compared to \BtoDpi decays. A single CB PDF is sufficient to describe this shape, which is fixed from fits to the simulation weighted by the PID efficiencies determined from calibration data. The same shapes are used for both \D decays, but are separately determined for the two \KS track types. The misidentified yields are determined in the global fit separately for each category.

The background observed at invariant masses to the left of the signal peak are candidates that originate from other \B-meson decays where not all of the decay products have been reconstructed. This background type is split into three sources: the first where the candidate originates from a \Bpm or \Bz meson where a photon or pion has not been reconstructed, referred to as partially reconstructed background; the second where the candidate originates from a \Bs meson; and the third where the candidate originates from a \Bpm or \Bz and one of the reconstructed tracks is assigned the kaon hypothesis when the true particle is a pion. The latter is referred to as misidentified partially reconstructed background and is only relevant to the \BtoDK category. The corresponding background in the \BtoDpi candidates is negligible due to the lower branching fractions of the background, and suppression due to the lower invariant-mass requirement. 

The first type of background has an invariant-mass distribution that can be determined analytically from the spin and mass of the missing particle~\cite{LHCb-PAPER-2017-021} convolved with a resolution function. The backgrounds that appear in the \BtoDpi category are: decays of $\decay{\Bpm}{\Dstarz \pipm}$ where the photon or pion from the $\Dstarz$ decay is not reconstructed, $\decay{B^0}{\Dstarpm \pimp}$ decays where the charged pion from the \Dstarpm decay is not reconstructed, and $\decay{\B^{0(\pm)}}{\D \pipm\pi^{\mp(0)}}$ decays where one companion pion is not reconstructed. The parameters of the shapes for the partially reconstructed backgrounds are determined through fits to simulation where elements of the resolution are shared across different decays. The main width parameter, which is the mass resolution of the $\decay{\Bpm}{\Dstarz\pipm}$ decay with a missing $\piz$, is determined through the fit to data that accounts for differences between simulation and data. 
The widths of other partially reconstructed background decays are fixed as the product of the main width parameter and the corresponding width ratios relative to $\decay{\Bpm}{\Dstarz\pipm}$, with these ratios determined from simulation.

In the \BtoDK category the partially reconstructed background comes from the counterpart decays $\decay{\B^\pm}{\Dstarz\Kpm}$, $\decay{B^0}{\Dstarpm \Kmp}$ and $\decay{B^{0(\pm)}}{\D \Kpm\pi^{\mp(0)}}$, where the shapes are also determined from simulation. 
The partially reconstructed background shapes describing processes where a particle species has been misidentified are also obtained from simulation, weighted by efficiencies determined from calibration data. In the global fit all parameters are fixed from simulation except for a mass shift that is shared for all components, and the main width of the resolution function which is different between the two \KS categories. 

The $\Bs \to \Dzb \Km \pip$ is a background that can contaminate the \BtoDK category close to the left of the signal peak when the pion is not reconstructed. This background is not visible in Figs.~\ref{fig:global_fit_res1}~and~\ref{fig:global_fit_res2} but is particularly problematic due to \Dzb contamination in a sample that will contain predominantly \Dz mesons. This means that the background is relatively large in the phase-space regions where the \DtoKshh decay process is suppressed. The shape of this background is determined from simulation that is weighted to match the amplitude model of the decay measured in Ref.~\cite{LHCb-PAPER-2014-036}. The shape is then determined by fitting to a mixture of the analytic functions used for the partially reconstructed background, including the shared mass shift which varies in the fit to data. The $B^0 \to D \Km \pip$ shape is determined in a similar way using the amplitude model determined in Ref.~\cite{LHCb-PAPER-2015-017}. The same distribution is also used for $\Bp \to D \Kp \piz$ decays. 

The combinatorial background component is modelled by an exponential PDF and is allowed to have a freely floating yield and slope parameter in each reconstruction category. The ratio of $\Dstar\to D\gamma$ to $\Dstar\to D\piz$ decays in  $\B \to \Dstar \pi$ backgrounds is determined in the fit, as is the ratio between the $\B\to\D\pi\pi$ and $B\to D^*\pi$ decays. For the other backgrounds, the relative fraction is fixed with respect to the \BtoDpi signal based on the ratio of branching fractions and efficiencies.

The fit method is validated using pseudoexperiments generated according to the baseline-fit result. The central values and uncertainties of the parameters determined in the global mass fit are confirmed to be unbiased. 
The fitted yields of \BtoDpi are about 17$\%$ higher than those in Ref.~\cite{LHCb-PAPER-2020-019}. Taking into account the change in integrated luminosity, there is an improvement in signal selection efficiency of a factor of around 2.7 for the \emph{long} \KS category, which is due to the new triggering system and the ability to select lower momentum tracks. 
Approximately 55$\%$ of the \BtoDpi signal in the 2024 data is found in the \emph{long} category in contrast to 30$\%$ in Run 1 and Run 2. This is due to a difference in the detector and reconstruction performance.  The yields of the different signals and background types are computed by integrating the PDFs in the region [5249, 5309]\mevcc to highlight the contributions near the signal peak and are reported in Table~\ref{tab:fit_yields}. The \BtoDK yields in categories of different \D decay and type of \KS candidate have uncertainties that are smaller than their Poisson uncertainty since they are determined using the value of ${R_{DK/D\pi}}$, which is measured from all \BtoDK candidates. The overall purity in the signal region is similar to the analysis in Ref.~\cite{LHCb-PAPER-2020-019}, although this analysis has less misidentified background and more combinatorial background. From the ratio of misidentified yields and signal yields obtained in the global mass fit, it is possible to determine the PID efficiencies for these data, which are inputs for the next stage of analysis. The probability that a companion pion passes the kaon hypothesis is approximately 1$\%$, and the probability that a companion kaon passes the pion hypothesis is approximately 27$\%$. These efficiencies are in agreement with those obtained from the calibration data.
\begin{table}[]
    \centering
        \caption{Signal and background yields in the invariant-mass range [5249, 5309]\mevcc as obtained in the fit. For the \BtoDK candidates, the yield of the partially reconstructed background includes the contributions from \Bs decays and misidentified partially reconstructed backgrounds. }
    \label{tab:fit_yields}
    \footnotesize
\begin{tabular}{ll|rr|rr}
\hline
\ &  & \multicolumn{2}{c|}{$\Bpm\to D\Kpm$ category} & \multicolumn{2}{c}{$\Bpm\to D\pipm$ category} \\
\hline
$D$ decay & Component & \multicolumn{1}{c}{\emph{long}} & \multicolumn{1}{c|}{\emph{downstream}} & \multicolumn{1}{c}{\emph{long}} & \multicolumn{1}{c}{\emph{downstream}} \\
\hline
$D\to K_{\rm S}^{0}\pip\pim$ & $\Bpm\to D\Kpm$        & $7545 \pm 74$          & $6203 \pm 61$          & $833 \pm \phantom{0}75$ & $471 \pm \phantom{0}42$ \\
                             & $\Bpm\to D\pipm$       & $161 \pm 10$           & $106 \pm \phantom{0}7$ & $113\,472 \pm 377$      & $92\,952 \pm 328$       \\
                             & Part. reco. background & $128 \pm \phantom{0}1$ & $129 \pm \phantom{0}1$ & $33 \pm \phantom{00}1$  & $27 \pm \phantom{00}1$  \\
                             & Combinatorial          & $547 \pm 18$           & $518 \pm 18$           & $845 \pm \phantom{0}23$ & $871 \pm \phantom{0}24$ \\
\hline    
$D\to K_{\rm S}^{0}\Kp\Km$   & $\Bpm\to D\Kpm$        & $1434 \pm 17$          & $1021 \pm 13$         & $128 \pm \phantom{0}30$  & $81 \pm \phantom{0}17$  \\
                             & $\Bpm\to D\pipm$       & $26 \pm \phantom{0}5$  & $21 \pm \phantom{0}3$ & $21\,409 \pm 152$        & $15\,042 \pm 124$       \\
                             & Part. reco. background & $24 \pm \phantom{0}1$  & $20 \pm \phantom{0}1$ & $6 \pm \phantom{00}1$    & $4 \pm \phantom{00}1$   \\
                             & Combinatorial          & $156 \pm \phantom{0}8$ & $98 \pm \phantom{0}7$ & $623 \pm \phantom{0}19$  & $334 \pm \phantom{0}15$ \\
\hline
\end{tabular}

\end{table}

\section {Determination of the \boldmath{\CP} parameters}
\label{sec:cpfit}

The data are subsequently divided into 160 subsets which come from the two $\B$ decays, two $B$ meson charges, two types of \KS reconstruction, $2\times8$ Dalitz-plot bins in the \DtoKspp decay, and $2\times 2$ Dalitz-plot bins in the \DtoKskk decay. A simultaneous fit to the invariant-mass distribution in the 160 subsets is performed to determine the values of the six \CP parameters and is denoted as the $\CP$ fit.  The lower limit of the invariant mass is increased to 5150 \mevcc to remove a large fraction of the partially reconstructed background. The shape parameters for each subset in the $\CP$ fit are fixed from the relevant global mass fit category described in Sect.~\ref{sec:massfit} and are different depending on the \B decay, \D decay and \KS reconstruction type. 

The signal yield in each subset does not vary independently. Instead, the signal yield is parametrised using Eq.~\eqref{eq:populations} or the analogous set of expressions for \BtoDpi. The parameters \xpmdk, \ypmdk are free parameters in the fit and are determined from the subsets with the relevant \B meson charge. The parameters \xxidpi, and \yxidpi are determined from the subsets in the \BtoDpi decay category. The parameters \ci and \si are fixed to the measured values discussed in Sect.~\ref{sec:theory}. Due to the different selection-efficiency variations over the Dalitz plot for the \emph{long} and \emph{downstream} \KS reconstruction categories, two sets of $F_i$ parameters (as defined in Eq.~\eqref{eq:fi}) are determined in the fit for each \D decay: one for each \KS category.  The constraints $\sum_i F_i = 1,\,\Fi\in[0,1]$ are incorporated into the fit using the formulation given in Ref.~\cite{LHCb-PAPER-2020-019}. The signal yield equations are normalised such that the parameters $h^{X}_{\Bpm}$, defined in Eq.~\eqref{eq:populations}, represent the total observed signal yield for subsets with the same \B decay, \B charge, \D decay and \KS reconstruction type, and these are also determined by the fit. 

The yield of combinatorial background in each subset is a free parameter. The yield of the total partially reconstructed background originating from \Bpm and \Bz decays is also a free parameter in each subset, where the shape is the sum of the relevant PDFs determined from the global mass fit. Relative yields between different background processes are recalculated to take into account the change in the lower invariant-mass limit. The yields of the misidentified background in each subset are determined from the product of the yield of the correctly-identified signal in the corresponding subset in the other $B$ decay and the PID efficiency extracted from the global mass fit. In the \BtoDK category, the misidentified partially reconstructed background yield in each subset is fixed by the partially reconstructed background in the corresponding \BtoDpi subset and the value of the PID efficiency. The total yield of the \Bs background is determined for each \D decay and \KS reconstruction type by the global mass fit. The yield of the \Bs background in each subset is therefore the total yield split evenly in $B$ charge and distributed across the Dalitz plot proportional to $F_{-i}$ ($F_{+i}$) in the \Bm (\Bp) decays, which respects the expected flavour of the \D decay. In some of the Dalitz-plot regions the yield of combinatorial or partially reconstructed background can be very low and subsequently lead to fit convergence issues. Consequently, if any yields of these backgrounds are found to be smaller than one, they are fixed to zero and the fit is repeated. The robustness of the fit procedure is evaluated using pseudoexperiments, including the modelling of backgrounds with low yields. The pseudoexperiments indicate that the uncertainties on the \CP observables are well estimated. A small bias in the central values is observed and assigned as a systematic uncertainty. This bias is apparent when signal and background yields are at baseline fit values and decreases as these yields are increased in the pseudoexperiments. 

\begin{figure}[tb]
    \centering
    \includegraphics[width=\linewidth]{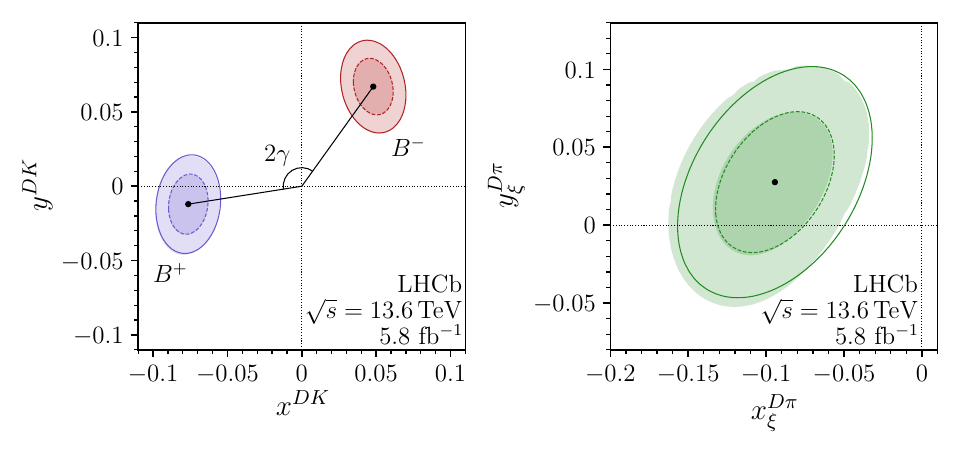}
    \caption{Confidence levels at 68.3\,\% and 95.4\,\% probability for (left, blue) $(\xp^{DK},\yp^{DK})$, (left, red) $(\xm^{DK}, \ym^{DK})$, and (right, green) $(x_\xi^{\D\pi},y_\xi^{\D\pi}) $ as measured in \BtoDK and \BtoDpi decays. The shaded regions are the result of a profile likelihood scan. The dashed and solid lines are ellipses drawn using the covariance from the fit result. The black dots show the central values.}
    \label{fig:ld_scan}
\end{figure}

The results are presented in Fig.~\ref{fig:ld_scan}. The agreement between the profile likelihood regions and those determined by the fit show that the uncertainties are close to a Gaussian approximation. Geometrically, the vectors to the points (\xpdk,\ypdk), (\xmdk,\ymdk) subtend an angle of 2\g. 

\begin{figure}[tb]
    \centering
    \includegraphics[width=0.9\linewidth]{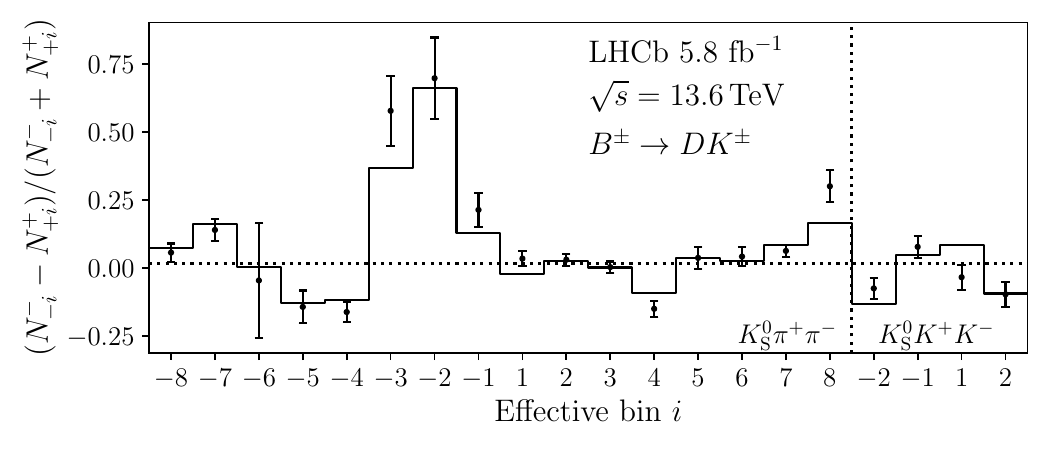}
    \includegraphics[width=0.9\linewidth]{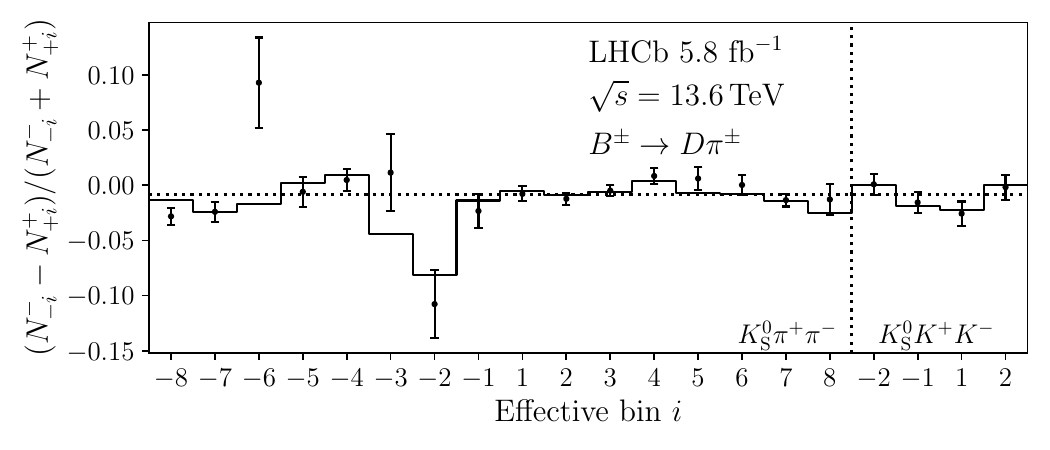}
    \caption{Bin-by-bin asymmetries $(N^-_{-i}-N^+_{+i})/(N^-_{-i}+N^+_{+i})$ for each Dalitz-plot bin number for (top) \BtoDK decays and (bottom) \BtoDpi decays. The prediction from the central values of the \CP-violation observables is shown with a solid line and the asymmetries obtained in fits with independent bin yields are shown with the error bars. The horizontal dashed lines represent average asymmetries over the phase space, while the vertical dashed lines separate the \Kspipi and \KsKK bins on the horizontal axes.}
    \label{fig:assym}
\end{figure}

An alternative fit is performed, where the signal yield of the correctly reconstructed \BtoDh decay is a floating parameter in each subset, to validate the robustness of the fit results. In this fit the \CP observables and \Fi parameters are not determined. From this alternative fit, asymmetries are computed for \emph{effective bin pairs}, defined to comprise bin $+i$ for a \Bp decay and bin $-i$ for a \Bm decay, and are shown in Fig.~\ref{fig:assym}. These results show that both the magnitude and sign of the asymmetry vary across the phase space, and that, as expected, the asymmetries are significantly larger for \BtoDK than for \BtoDpi decays. The larger uncertainties in several bins arise from limited signal statistics in these phase-space regions. The results of the alternative fit can be compared with those of the baseline fit. The expected asymmetries, calculated from the total signal yields and \CP observables, are shown as solid lines in Fig.~\ref{fig:assym}. The two visualisations of the asymmetry in different bins of the Dalitz plot are naturally correlated since they are determined from the same data. The expected spread between the data points and the solid line in Fig.~\ref{fig:assym} is investigated using pseudoexperiments. These confirm that the differences between the two methods of visualising the asymmetries in data are consistent within expected statistical fluctuations of the two methods. This also implies that Eq.~\eqref{eq:populations} provides a good description of the data. 

\section{Systematic uncertainties}
\label{sec:syst}
The sources of systematic uncertainty and their sizes for the \CP observables are summarised in Table~\ref{tab:systematics}. Two different methods are used for determining the size of the systematic uncertainty depending on its origin. One method resamples a parameter within its uncertainty, refits to data and assigns the width of the resulting \CP observables as the systematic uncertainty. Correlations between \CP observables are determined from the Pearson correlation coefficient obtained using each pair of the fitted parameters. The other method generates pseudoexperiments in an alternative configuration and fits using the baseline fit model. In this case, the difference between the mean of the baseline and the alternate configurations is assigned as the systematic uncertainty, and correlations between the \CP observables are assigned $\pm 1$ depending on the sign of the bias.
The \hyperref[sec:app]{Appendix} provides the correlation matrix for the strong-phase uncertainty coming from the external parameters and the summed correlation matrix for all other systematic uncertainties.

The uncertainties on the strong-phase inputs, \ci and \si, are propagated to the \CP observables by resampling their fixed values according to the covariance matrices in Refs.~\cite{BESIII:2025nsp} and \cite{Krishna}, repeating the fit for each sample, and computing the covariance of the resulting physics parameters. The size of the uncertainty is found to have some dependence on the $B$ data sample itself~\cite{BELLEMODIND} as well as the central values and uncertainties of \ci and \si and the values of the \CP observables. The size of the observed uncertainty is checked by performing the same study on pseudoexperiments and it is found that the determined uncertainties on data are within the range found in pseudoexperiments. 

A significant source of systematic uncertainty arises from the assumption that there is no variation in the component PDFs used to describe the invariant-mass distributions in different Dalitz-plot bins. The dominant effect is caused by the assumption for the combinatorial background. The slope of the combinatorial background is examined in data from the upper \Bpm mass sideband in each Dalitz-plot bin and compared to the slope from the global mass fit. Pseudoexperiments are generated with these differences and fitted with the assumption of uniform slopes in all Dalitz-plot bins. The mean biases in the \CP parameters are taken as the systematic uncertainties. This source of uncertainty is determined to be larger than the corresponding uncertainty in Ref.~\cite{LHCb-PAPER-2020-019}, due to the combination of the larger level of combinatorial background, and larger variation across the phase space found in the data collected in 2024. The larger number of $pp$ collisions per bunch crossing in Run~3 compared to Run~2 leads to a higher combinatorial rate. Other PDF components are also investigated for changes across the Dalitz plot. Simulation is used to determine the difference of the signal and misidentified mass resolutions in different Dalitz-plot bins. The impact of these differences is also investigated using pseudoexperiments and results in a small contribution to the uncertainty. 

The assumption that the partially reconstructed background shape is the same across different Dalitz-plot bins is also investigated as a source of systematic uncertainty. In this case, the difference is driven by individual components that make up the total shape having differing distributions over the Dalitz plot. To assess the impact of this assumption, pseudoexperiments are generated based on alternative background models which allow for variation across the Dalitz plot and fit back with the baseline model. The partially reconstructed backgrounds will contain a mixture of \Dz and \Dzb decays, where the mixture is determined by the underlying physics for each component decay and the particle that is missed. For $B \to D^{(*)}K^{(*)}$ backgrounds, where \CP violation is present, the different contributions for each bin are determined using the hadronic parameters, $r_B^{D^{(*)}K^{(*)}}$ and $\delta_B^{D^{(*)}K^{(*)}}$ from Ref.~\cite{LHCb-CONF-2024-004}. The corresponding hadronic parameters for $\Bpm \to D \Kstarpm$ and $\Bz \to D\Kstarz$ are used as a proxy for the contributions from $\Bz \to D\Kpm\pimp$ and $\Bpm \to D \Kpm \piz$ decays, respectively. No variation is introduced to the partially reconstructed backgrounds in the \BtoDpi category since any \CP violation is expected to be very small. The mean bias in each \CP parameter from the pseudoexperiments is assigned as the systematic uncertainty. 

The other large contribution to the systematic uncertainties comes from not modelling some small backgrounds in the fit, which all have estimated yields less than 1$\%$ of the signal yield. The largest of these originates from misidentified $\Lb \to D p \pim$ decays in the \BtoDK sample, where the charged pion is not reconstructed and the proton is misidentified as a kaon,  which contain the opposite flavour $D$ meson to the majority of the signal. Hence the relative fraction of this background is enhanced in regions of the Dalitz plot where the signal is suppressed, leading to the dominant source of this systematic uncertainty. Other backgrounds that are also ignored are decays of $\Bp \to D \mup \nu_\mu$ and $\Bp\to \Dzb h^+, \Dzb\to \Kstarp l^- \bar{\nu_l}$ where $l=e,\mu$. To estimate the size of the systematic uncertainty, the shapes of these backgrounds are determined from a fast parametric simulation~\cite{Cowan:2016tnm} and a contribution from them based on expected yields is inserted into the pseudoexperiment generation. The mean bias on the \CP parameters from fitting these ensembles with the baseline fit is taken as the systematic uncertainty.  There also remains some residual background that has the same final state as the signal but does not proceed via an intermediate charmed meson. The systematic uncertainty is determined from generating pseudoexperiments containing this component and then ignoring them in the fit and using the mean biases on the \CP parameters as the uncertainty. The parameters for the generation are determined from studies of the upper sideband of the \D-candidate invarant mass, where the rates of this background are determined to be approximately $0.5\%$ of the \BtoDK signal, and an order of magnitude smaller for the \BtoDpi channel. The distribution over the Dalitz plot is determined from fits of this background in the \D-candidate invariant-mass sideband with the data separated into the Dalitz-plot regions.  

The parameters of the mass shapes are fixed in the second-stage fit to determine the \CP observables and thus the uncertainties from the first-stage fit must be propagated forward. A resampling method as described in Ref.~\cite{LHCb-PAPER-2020-019} is used to determine this uncertainty. The fixed fractions in the fit are dependent on various branching fractions and selection efficiency ratios determined from simulation,  both of which have corresponding uncertainties. The impact of the fixed fractions are determined by resampling their values according to their uncertainties in the global fit and then reperforming the \CP fit with the new mass shape parameters. The covariance for this systematic uncertainty is determined from the distribution of \CP observables. A similar procedure is carried out for the PID efficiencies.  

The main effect of migration from one Dalitz-plot bin to another is implicitly taken into account by using data to determine the $F_i$ values, which thus includes the effects of net bin migration. However, a residual effect arises because of the small differences in the  net migrations of the \BtoDK and \BtoDpi decays due to the differing hadronic decay parameters. To investigate this, pseudoexperiments are generated according to the amplitude model in Ref.~\cite{BaBar:2018agf,Belle2018} with \CP observables from Ref.~\cite{LHCb-CONF-2025-003}. The Dalitz-plot position of the generated data are smeared based on the \lhcb simulation taking into account that the resolution on the Dalitz plot changes depending on the position within the Dalitz plot. The ensembles are fit with the baseline procedure, although in this case it is appropriate to fix the values of \ci and \si to those computed from the amplitude model rather than the measured values from BESIII. The mean bias in the \CP violation parameters is taken as the systematic uncertainty, which is relatively small compared to all other uncertainties of the analysis. 

In Sect.~\ref{sec:theory}, the effect of a nonuniform efficiency profile over the Dalitz plot is not introduced into the definition of \ci and \si. To investigate this impact, simulated samples based on amplitude models from Refs.~\cite{BaBar:2018agf,Belle2018} and\cite{ BABAR2010} for the \DtoKspp and \DtoKskk decays are used to determine the variation of the \ci and \si values due to this effect. Pseudoexperiments are then generated with adjusted values of \ci and \si and fit with the model values. The resulting bias is taken as the systematic uncertainty. 

Systematic effects due to neglecting \CP violation, mixing and regeneration in the neutral kaon system, and charm mixing are investigated by generating pseudoexperiments that contain these effects~\cite{KsCPV,BPV} and fitting the data using the baseline procedure. These effects along with the fit bias discussed in Sect.~\ref{sec:cpfit} are small compared to the other uncertainties.

\begin{table}[t]
    \centering
    \caption{Summary of the uncertainties on the \CP observables. All numbers are absolute uncertainties quoted in units of $10^{-2}$. }
    \label{tab:systematics}
    \newcolumntype{R}[1]{>{\raggedleft\let\newline\\\arraybackslash\hspace{0pt}}m{#1}}
    \begin{tabular}{l|R{0.08\linewidth}R{0.08\linewidth}R{0.08\linewidth}R{0.08\linewidth}R{0.08\linewidth}R{0.08\linewidth}}
    \hline \\[-1em]
    \multicolumn{1}{r|}{Uncertainty on} &  $x_-^{DK}$ & $y_-^{DK}$ & $x_+^{DK}$ & $y_+^{DK}$ & $x_{\xi}^{D\pi}$ & $y_{\xi}^{D\pi}$ \\ \\[-1em]
    \hline
    Statistical                  & 0.88 & 1.25 & 0.87 & 1.30 & 2.55 & 3.05 \\ 
    \hline
    Strong-phase inputs \cite{BESIII:2025nsp}         & 0.23 & 0.56 & 0.15 & 0.44 & 0.69 & 1.21 \\
    \hline
    Total \lhcb systematic  & 0.20 & 0.44 & 0.28 & 0.35 & 0.57 & 0.19 \\ \\[-1em]
 \phantom{xx}\PDF in Dalitz-plot bins     & 0.08 & 0.28 & 0.22 & 0.25 & 0.41 & 0.01 \\
\phantom{xx}Part. reco. background       & 0.10 & 0.04 & 0.04 & 0.05 & 0.16 & 0.01 \\
    
  \phantom{xx}Small backgrounds             & 0.09 & 0.25 & 0.13 & 0.16 & 0.25 & 0.12 \\
   \phantom{xx}Mass shape parameters         & 0.03 & 0.06 & 0.07 & 0.03 & 0.22 & 0.04 \\
    \phantom{xx}Fixed yield ratios            & 0.05 & 0.11 & 0.05 & 0.05 & 0.03 & 0.04 \\
    \phantom{xx}PID efficiencies              & 0.03 & 0.08 & 0.03 & 0.03 & 0.02 & 0.02 \\

    \phantom{xx}Migration  & 0.01 & 0.14 & 0.02 & $<0.01$ & 0.07 & 0.03 \\
    
   \phantom{xx}Efficiency effects on $c_i, s_i$ & 0.08 & 0.09 & $<0.01$ & 0.13 & 0.06 & 0.10 \\
    \phantom{xx}$\KS$ effects & $<0.01$ & 0.03 & 0.07 & 0.10 & 0.06 & 0.07 \\
    \phantom{xx}$D$ mixing                    & 0.04 & 0.02 & 0.01 & 0.03 & 0.01 & 0.02 \\
    \phantom{xx}Fit bias                   & 0.03 & 0.03 & 0.02 & 0.02 & 0.08 & 0.04 \\
    \hline
    \end{tabular}
\end{table}

Beyond the evaluation of the systematic uncertainties described above, a series of cross checks are also carried out. Fits are performed to separate samples where the data sample is split into smaller data-taking periods, type of \KS candidate, \D decay, and magnet polarity. The results are consistent between the different samples. Further studies are carried out where the data are split as a function of the $B$ meson momentum, the $B$ meson transverse momentum, and the pseudorapidity of the $B$ meson. Other splits divide the data by the number of primary vertices in the event or the value of the BDT variable. Comparisons are made of the \CP observables, and also the interpreted physics parameters \g, \rBDK, \rBDpi, \dBDK, and \dBDpi. The statistical correlation between the combined results across a split category and the baseline fit is not $100\%$ due to changes in purity in different subsamples which change the per-event statistical weight. Therefore the expected shifts in the comparison are determined from pseudoexperiments that model the splitting of the data. All shifts are found to be consistent with expectations and no significant trends are found in the data.

\section{Results and interpretation}
\label{sec:interpretation}

The \CP observables are measured to be
\begin{align*}
        \xmdk   &= (\phantom{-}4.81 \pm 0.88 \pm 0.20 \pm 0.23) \times 10^{-2}, \\
        \ymdk   &= (\phantom{-}6.70 \pm 1.26 \pm 0.44 \pm 0.56) \times 10^{-2}, \\
        \xpdk   &= (-7.63 \pm 0.88 \pm 0.28 \pm 0.15) \times 10^{-2}, \\
        \ypdk   &= (-1.20 \pm 1.34 \pm 0.35 \pm 0.44) \times 10^{-2}, \\
        \xxidpi &= (-9.44 \pm 2.51 \pm 0.57 \pm 0.69) \times 10^{-2}, \\
        \yxidpi &= (\phantom{-}2.76 \pm 2.99 \pm 0.19 \pm 1.21) \times 10^{-2},
\end{align*} 
where the first uncertainty is statistical, the second systematic, and the third from strong-phase inputs. The correlation matrices for each source of uncertainty are provided in the Appendix. The determined central values, uncertainties, and correlations are reported in an associated HEPData record~\cite{hepdata}.  The results are consistent with the measurements from Ref.~\cite{LHCb-PAPER-2020-019}. 
They can be interpreted in terms of five parameters, $(\g, \rBDK, \dBDK, \rBDpi, \dBDpi)$ according to
\begin{equation}
    \begin{array}{r@{}l}
        \xpmdk + i\ypmdk &= \rBDK\exp(i(\dBDK\pm\g)), \\
        \xxidpi + i\yxidpi &= \displaystyle\frac{\rBDpi}{\rBDK}\exp(i(\dBDpi - \dBDK)).
    \end{array}
\end{equation}
The parameters \g, \rB and \dB for each \Bpm decay are determined via a maximum likelihood fit using a frequentist treatment (\textsc{Plugin}) as described in Refs.~\cite{LHCb-PAPER-2016-032,LHCb-PAPER-2021-033}. Small biases in \rBDK, \rBDpi and \dBDpi are observed in pseudoexperiments. Studies confirm these biases originate from the sample size since they decrease when the signal and background yields are increased. The interpretation is corrected for these biases which are determined using pseudoexperiments. There is a two-fold ambiguity in the results due to the trigonometric relations where the results are invariant with the addition of 180$^\circ$ to both \g and \dB. The solution that satisfies $0 < \g < 180^\circ$ is chosen, and leads to 
\begin{align*}
        \gamma & = (68.1\pm 6.7)^{\circ}, \\
         \rBDK & = 0.0781^{+0.0078}_{-0.0079}, \\
         \dBDK & = (121.5^{+6.9}_{-7.4})^{\circ}, \\
        \rBDpi & = 0.0073^{+0.0016}_{-0.0015}, \\
        \dBDpi & = (286^{+20}_{-23})^{\circ},
\end{align*}
where the uncertainties include both statistical and systematic contributions.

\begin{figure}[tb]
    \centering
     \includegraphics[width=0.49\linewidth]{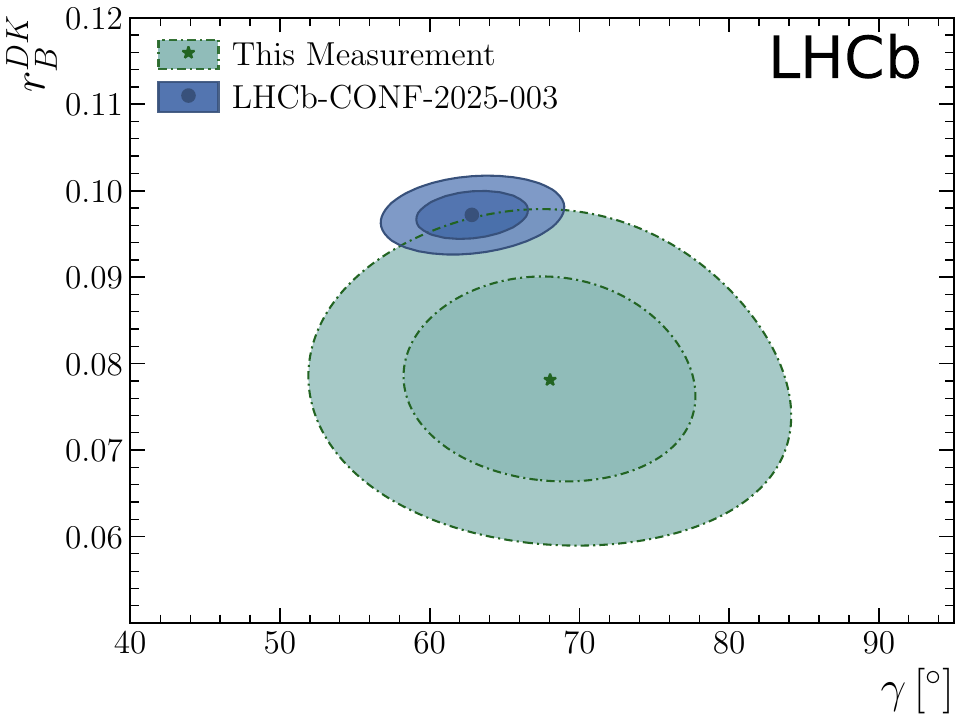}
     \includegraphics[width=0.49\linewidth]{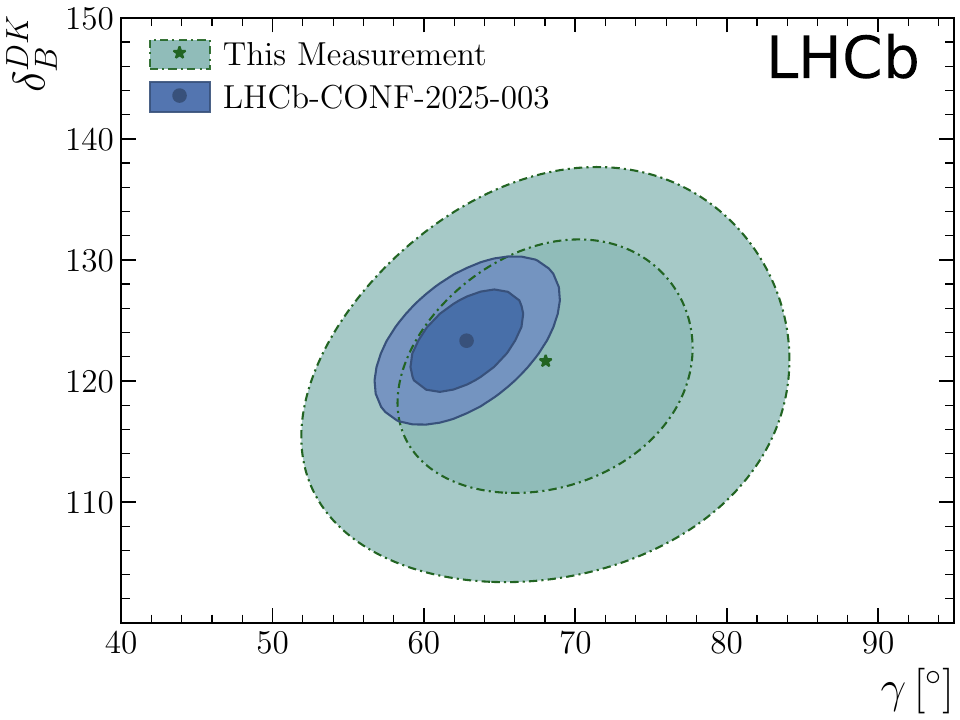} 
    \caption{Two-dimensional confidence level contours in the (left) $\gamma$ \vs $\rBDK$ and (right) $\gamma$ \vs $\dBDK$ planes for this measurement compared to that presented in Ref.\cite{LHCb-CONF-2025-003} using the \textsc{Prob} method described in Ref.~\cite{LHCb-PAPER-2016-032}. The inner and outer contours contain 68.3\% and 95.4\% of the likelihood distribution, respectively.}
    \label{fig:2d_scan}
\end{figure}

A comparison of these results and those in Ref.~\cite{LHCb-CONF-2025-003} yields a $p$-value of 12$\%$ when considering the full five-dimensional parameter space. Representations of the comparison are shown in Fig.~\ref{fig:2d_scan} where good agreement is seen in \g and \dBDK, and a small but not statistically significant tension is observed in \rBDK.  It is noted that the uncertainty on \g is around 20$\%$ larger in this measurement than that presented in Ref.~\cite{LHCb-PAPER-2020-019} despite the signal yields being higher. The lower sensitivity arises due to two effects. The first is due to the change in central values of the \ci and \si parameters of the \DtoKspp decay measured with a larger dataset and improved methodology~\cite{BESIII:2025nsp}. It is verified that the per-event sensitivity is reduced with the new central values using pseudoexperiments. The second is due to the lower determined value of \rBDK, which scales inversely with the uncertainty of~\g.

\section{Conclusions}
\label{sec:conclusions}

The CKM angle \g is measured with the decays \BtoDK and \BtoDpi  with \DtoKspp or ${\DtoKsKK}$, using 5.8\invfb of data collected by the \lhcb experiment in 2024.
 The sensitivity to \g comes almost entirely from \BtoDK decays where the signal yields of reconstructed events are approximately 14\,000 (2500) in the \DtoKspp (\DtoKskk) decay modes. The \BtoDpi data are primarily used to control effects due to selection and reconstruction of the data. This is of particular importance and leads to minimal dependence on simulation and insensitivity to changes in the data taking conditions. The analysis is performed in bins of the \D-decay Dalitz plot and measurements of the strong-phase parameters (\ci, \si) from \besiii for \DtoKspp~\cite{BESIII:2025nsp} and \besiii and \cleo for \DtoKskk~\cite{Krishna} are used to provide essential input. Such an approach allows the analysis to be free from model-dependent assumptions on the strong-phase variation across the Dalitz plot. The analysis also determines the hadronic parameters \rB and \dB for each \Bpm decay mode. The CKM angle \g is determined to be $\g = (68.1\pm6.7)^\circ$, where the precision is limited by the statistical uncertainty.

\section*{Acknowledgements}
%
%
\noindent We express our gratitude to our colleagues in the CERN
accelerator departments for the excellent performance of the LHC. We
thank the technical and administrative staff at the LHCb
institutes.
We acknowledge support from CERN and from the national agencies:
ARC (Australia);
CAPES, CNPq, FAPERJ and FINEP (Brazil); 
MOST and NSFC (China); 
CNRS/IN2P3 and CEA (France);  
BMFTR, DFG and MPG (Germany);
INFN (Italy); 
NWO (Netherlands); 
MNiSW and NCN (Poland); 
MEC/IFA (Romania); 
MICIU and AEI (Spain);
SNSF and SER (Switzerland); 
NASU (Ukraine); 
STFC (United Kingdom); 
DOE NP and NSF (USA).
We acknowledge the computing resources that are provided by ARDC (Australia), 
CBPF (Brazil),
CERN, 
IHEP and LZU (China),
IN2P3 (France), 
KIT and DESY (Germany), 
INFN (Italy), 
SURF (Netherlands),
Polish WLCG (Poland),
IFIN-HH (Romania), 
PIC (Spain), CSCS (Switzerland), 
GridPP (United Kingdom),
and NSF (USA).  
We are indebted to the communities behind the multiple open-source
software packages on which we depend.
Individual groups or members have received support from
RTP (Australia), 
FWO Odysseus grant G0ASD25N (Belgium), 
Key Research Program of Frontier Sciences of CAS, CAS PIFI, CAS CCEPP (China); 
Minciencias (Colombia);
EPLANET, Marie Sk\l{}odowska-Curie Actions, ERC and NextGenerationEU (European Union);
A*MIDEX, ANR, IPhU and Labex P2IO, and R\'{e}gion Auvergne-Rh\^{o}ne-Alpes (France);
Alexander-von-Humboldt Foundation (Germany);
ICSC (Italy); 
Severo Ochoa and Mar\'ia de Maeztu Units of Excellence, GVA, XuntaGal, GENCAT, InTalent-Inditex and Prog.~Atracci\'on Talento CM (Spain);
the Leverhulme Trust, the Royal Society and UKRI (United Kingdom).

\clearpage
\appendix
\section*{Appendix}
\label{sec:app}

The correlations matrices for the measured observables are shown in Tables~\ref{tab:cpfit_res_xy_stat_corr}--\ref{tab:strong_phase_uncertainty} for the statistical uncertainties, the experimental systematic uncertainties, and the strong-phase-related uncertainties, respectively.
\begin{table}[h]
    \caption{Correlation matrix of the statistical uncertainties of \CP observables.}
    \label{tab:cpfit_res_xy_stat_corr}
    \renewcommand{\arraystretch}{1.2}
    \centering
    \begin{tabular}{l|rrrrrr}
    
         & $x_-^{DK}$ & $y_-^{DK}$ & $x_+^{DK}$ &$ y_+^{DK}$ & $x_{\xi}^{D\pi}$ &$ y_{\xi}^{D\pi} $\\
        \hline
        $x_-^{DK}        $ & $\phantom{-}1.000$ & $-0.183$ & $-0.034$ & $\phantom{-}0.024$ & $\phantom{-}0.021$ & $-0.260$ \\
        $y_-^{DK}        $ & & $\phantom{-}1.000$ & $-0.041$ & $-0.052$ & $\phantom{-}0.233$ & $\phantom{-}0.164$ \\
        $x_+^{DK}        $ & & & $\phantom{-}1.000$ & $\phantom{-}0.105$ & $-0.241$ & $\phantom{-}0.025$ \\
        $y_+^{DK}        $ & & & & $\phantom{-}1.000$ & $-0.258$ & $-0.189$ \\
        $x_{\xi}^{D\pi}$   & & & & & $\phantom{-}1.000$ & $\phantom{-}0.381$ \\
        $y_{\xi}^{D\pi}$   & & & & & & $\phantom{-}1.000$ \\
        
    \end{tabular}
\end{table}

\begin{table}[h]
    \caption{Correlation matrix of the systematic uncertainties from the \lhcb experiment.}
    \label{tab:lhcb_syst_corr}
    \renewcommand{\arraystretch}{1.2}
    \centering
    \begin{tabular}{l|rrrrrr}
        
        & $x_-^{DK}$ & $y_-^{DK}$ & $x_+^{DK}$ &$ y_+^{DK}$ & $x_{\xi}^{D\pi}$ &$ y_{\xi}^{D\pi} $\\
        \hline
        $x_-^{DK}        $ & $\phantom{-}1.000$ & $\phantom{-}0.009$ & $\phantom{-}0.271$ & $-0.321$ & $\phantom{-}0.115$ & $-0.169$ \\
        $y_-^{DK}        $ &    & $\phantom{-}1.000$ & $-0.585$ & $\phantom{-}0.609$ & $\phantom{-}0.641$ & $-0.109$ \\
        $x_+^{DK}        $ & &       & $\phantom{-}1.000$ & $-0.353$ & $-0.730$ & $\phantom{-}0.322$ \\
        $y_+^{DK}        $ & & &          & $\phantom{-}1.000$ & $\phantom{-}0.500$ & $\phantom{-}0.283$ \\
        $x_{\xi}^{D\pi}$  & & &     &         & $\phantom{-}1.000$ & $\phantom{-}0.024$ \\
        $y_{\xi}^{D\pi}$  & & &     &    &         & $\phantom{-}1.000$ \\
    
    \end{tabular}
\end{table}
\begin{table}[tb]
    \caption{Correlation matrix of the strong-phase related systematic uncertainties.}
    \renewcommand{\arraystretch}{1.2}
    \centering
    \begin{tabular}{l|rrrrrr}
      
        & $x_-^{DK}$ & $y_-^{DK}$ & $x_+^{DK}$ &$ y_+^{DK}$ & $x_{\xi}^{D\pi}$ &$ y_{\xi}^{D\pi} $\\
        \hline
        $x_-^{DK}        $ & $\phantom{-}1.000$ & $-0.112$ & $-0.235$ & $\phantom{-}0.617$ & $-0.283$ & $-0.300$ \\
        $y_-^{DK}        $ & & $\phantom{-}1.000$ & $\phantom{-}0.049$ & $\phantom{-}0.054$ & $-0.028$ & $\phantom{-}0.036$ \\
        $x_+^{DK}        $ & & & $\phantom{-}1.000$ & $\phantom{-}0.127$ & $-0.572$ & $-0.364$ \\
        $y_+^{DK}        $ & & & & $\phantom{-}1.000$ & $-0.501$ & $-0.546$ \\
        $x_{\xi}^{D\pi}$ & & & & & $\phantom{-}1.000$ & $\phantom{-}0.898$ \\
        $y_{\xi}^{D\pi}$ & & & & & & $\phantom{-}1.000$ \\
    
    \end{tabular}
    \label{tab:strong_phase_uncertainty}
\end{table}

\clearpage

\addcontentsline{toc}{section}{References}
\bibliographystyle{LHCb/LHCb}
\bibliography{main,LHCb/standard,LHCb/LHCb-PAPER,LHCb/LHCb-CONF,LHCb/LHCb-DP,LHCb/LHCb-TDR}

\newpage
\centerline
{\large\bf LHCb collaboration}
\begin
{flushleft}
\small
R.~Aaij$^{38}$\lhcborcid{0000-0003-0533-1952},
M.~Abdelfatah$^{69}$,
A.S.W.~Abdelmotteleb$^{57}$\lhcborcid{0000-0001-7905-0542},
C.~Abellan~Beteta$^{51}$\lhcborcid{0009-0009-0869-6798},
F.~Abudin\'en$^{59}$\lhcborcid{0000-0002-6737-3528},
T.~Ackernley$^{61}$\lhcborcid{0000-0002-5951-3498},
A.A.~Adefisoye$^{69}$\lhcborcid{0000-0003-2448-1550},
B.~Adeva$^{47}$\lhcborcid{0000-0001-9756-3712},
M.~Adinolfi$^{55}$\lhcborcid{0000-0002-1326-1264},
P.~Adlarson$^{87,42}$\lhcborcid{0000-0001-6280-3851}
% ...

{\footnotesize \it

% ...
$^{38}$Nikhef National Institute for Subatomic Physics, Amsterdam, Netherlands\\
% ...
$^{42}$National Center for Nuclear Research (NCBJ), Warsaw, Poland\\
% ...
$^{47}$Instituto Galego de F{\'\i}sica de Altas Enerx{\'\i}as (IGFAE), Universidade de Santiago de Compostela, Santiago de Compostela, Spain\\
% ...
$^{51}$Physik-Institut, Universit{\"a}t Z{\"u}rich, Z{\"u}rich, Switzerland\\
% ...
$^{55}$H.H. Wills Physics Laboratory, University of Bristol, Bristol, United Kingdom\\
% ...
$^{57}$Department of Physics, University of Warwick, Coventry, United Kingdom\\
% ...
$^{59}$School of Physics and Astronomy, University of Edinburgh, Edinburgh, United Kingdom\\
$^{60}$School of Physics and Astronomy, University of Glasgow, Glasgow, United Kingdom\\
$^{61}$Oliver Lodge Laboratory, University of Liverpool, Liverpool, United Kingdom\\
% ...
$^{69}$Syracuse University, Syracuse, NY, United States\\
% ...
$^{87}$Department of Physics and Astronomy, Uppsala University, Uppsala, Sweden, associated to $^{60}$\\
% ...
}
\end{flushleft}

\end{document}